%% file: main.tex
\documentclass[a4paper,UKenglish,cleveref, autoref, thm-restate]{lipics-v2021}

\input{macro}

\title{Thinking Fast and Slow: Data-Driven Adaptive DeFi Borrow-Lending Protocol}
\titlerunning{Data-Driven Adaptive DeFi Borrow-Lending Protocol}

\author{Mahsa Bastankhah}
{Princeton University, Princeton, NJ, USA \and \url{https://sites.google.com/view/mahsabastankhah/home?authuser=0}}
{mhs.bastankhah@princeton.edu}
{}
{}

\author{Viraj Nadkarni}
{Princeton University, Princeton, NJ, USA  \and \url{https://sites.google.com/view/virajnadkarni/home}}
{vn7241@princeton.edu}
{}
{}

\author{Xuechao Wang}
{Hong Kong University of Science and Technology (Guangzhou), Guangdong, China  \and \url{https://xuechao2.github.io/}}
{xuechaowang@hkust-gz.edu.cn}
{}
{}

\author{Chi Jin}
{Princeton University, Princeton, NJ, USA  \and \url{https://sites.google.com/view/cjin/home}}
{chij@princeton.edu}
{}
{}

\author{Sanjeev Kulkarni}
{Princeton University, Princeton, NJ, USA  \and \url{https://www.princeton.edu/~kulkarni/}}
{kulkarni@princeton.edu}
{}
{}

\author{Pramod Viswanath}
{Princeton University, Princeton, NJ, USA  \and \url{https://ece.princeton.edu/people/pramod-viswanath}}
{pramodv@princeton.edu}
{}
{}

\authorrunning{M. Bastankhah, V. Nadkarni, X. Wang, C. Jin, S. Kulkarni, and P. Viswanath}

\Copyright{Mahsa Bastankhah, Viraj Nadkarni, Chi Jin, Sanjeev Kulkarni, and Pramod Viswanath}

\begin{document}
\maketitle

\begin{abstract}
Decentralized finance (DeFi) borrowing and lending platforms are crucial to the decentralized economy, involving two main participants: lenders who provide assets for interest and borrowers who offer collateral exceeding their debt and pay interest. Collateral volatility necessitates over-collateralization to protect lenders and ensure competitive returns. Traditional DeFi platforms use a fixed interest rate curve based on the utilization rate (the fraction of available assets borrowed) and determine over-collateralization offline through simulations to manage risk. This method doesn't adapt well to dynamic market changes, such as price fluctuations and evolving user needs, often resulting in losses for lenders or borrowers. In this paper, we introduce an adaptive, data-driven protocol for DeFi borrowing and lending. Our approach includes a high-frequency controller that dynamically adjusts interest rates to maintain market stability and competitiveness with external markets. Unlike traditional protocols, which rely on user reactions and often adjust slowly, our controller uses a learning-based algorithm to quickly find optimal interest rates, reducing the opportunity cost for users during periods of misalignment with external rates. Additionally, we use a low-frequency planner that analyzes user behavior to set an optimal over-collateralization ratio, balancing risk reduction with profit maximization over the long term. This dual approach is essential for adaptive markets: the short-term component maintains market stability, preventing exploitation, while the long-term planner optimizes market parameters to enhance profitability and reduce risks. We provide theoretical guarantees on the convergence rates and adversarial robustness of the short-term component and the long-term effectiveness of our protocol. Empirical validation confirms our protocol's theoretical benefits. 

\end{abstract}

\input{introduction}


\input{Problem_Formulation}

\input{protocol-theoretical}


\input{evaluation}

\input{system-design}

\input{conclusion}

\bibliography{references}  

\appendix

\input{appendix}

\end{document}

%% file: macro.tex
\nolinenumbers 

\usepackage{mathtools}
\usepackage{bbm}

\usepackage{bm}
\usepackage{xspace}
\usepackage{appendix} 
\usepackage{algorithm} 
\usepackage{algpseudocode} 
\usepackage{prettyref}
\newrefformat{fig}{Figure~[\ref{#1}]}
\newrefformat{app}{Appendix~[\ref{#1}]}
\newrefformat{eq}{Equation~[\ref{#1}]}

\newcommand{\indicator}[1]{\mathbbm{1}_{#1}}

\newcommand{\paramProt}{protocol\xspace}

\newcommand{\colfac}{collateral factor\xspace}
\newcommand{\liqthrsh}{liquidation threshold\xspace}

\newcommand{\colfacN}{\ensuremath{c}\xspace}
\newcommand{\ut}{\ensuremath{U_t}\xspace}
\newcommand{\uS}[1]{\ensuremath{U_{#1}}\xspace}
\newcommand{\rt}{\ensuremath{r_t}\xspace}
\newcommand{\rS}[1]{\ensuremath{r_{#1}}\xspace}
\newcommand{\liqthrshN}{\ensuremath{LT}\xspace}
\newcommand{\liqIncntive}{\ensuremath{LI}\xspace}
\newcommand{\lt}{\ensuremath{L_t}\xspace}
\newcommand{\lS}[1]{\ensuremath{L_{#1}}\xspace}

\newcommand{\rolt}{\ensuremath{r_{o}^l}\xspace}
\newcommand{\robt}{\ensuremath{r_{o}^b}\xspace}
\newcommand{\rolhat}{\ensuremath{\hat{r}_{o}^l}\xspace}
\newcommand{\robhat}{\ensuremath{\hat{r}_{o}^b}\xspace}
\newcommand{\rolbar}{\ensuremath{\bar{r}_{o}^l}\xspace}
\newcommand{\robbar}{\ensuremath{\bar{r}_{o}^b}\xspace}

\newcommand{\bt}{\ensuremath{B_t}\xspace}

\newcommand{\bS}[1]{\ensuremath{B_{#1}}\xspace}
\newcommand{\ct}{\ensuremath{C_t}\xspace}

\newcommand{\cS}[1]{\ensuremath{C_{#1}}\xspace}
\newcommand{\pt}{\ensuremath{p_t}\xspace}
\newcommand{\pS}[1]{\ensuremath{p_{#1}}\xspace}

\newcommand{\assetOne}{\ensuremath{\mathcal{A}_l}\xspace}
\newcommand{\assetTwo}{\ensuremath{\mathcal{A}_c}\xspace}

\newcommand{\userDefault}[2]{\ensuremath{\pi_{#1}^{#2}}\xspace}
\newcommand{\poolDefault}[1]{\ensuremath{\pi_{#1}}\xspace}
\newcommand{\userLiq}[2]{\ensuremath{\lambda_{#1}^{#2}}\xspace}

\newcommand{\Deltat}{\ensuremath{\Delta }\xspace}

\newcommand{\protocol}{\ensuremath{\mathcal{P}}\xspace}

\newcommand{\elasticityl}{\ensuremath{\eta_l}\xspace}
\newcommand{\elasticityb}{\ensuremath{\eta_b}\xspace}

\newcommand{\volat}{\ensuremath{\sigma}\xspace}
\newcommand{\muprice}{\ensuremath{\mu}\xspace}

\newcommand{\convrate}{\ensuremath{\mathcal{R}_{\protocol}(\delta,\tau)}\xspace}

\newcommand{\utilityl}{\ensuremath{\text{Utility}_t^l}\xspace}
\newcommand{\utilitybOne}{\ensuremath{\text{Utility}_t^{b,1}}\xspace}
\newcommand{\utilitybTwo}{\ensuremath{\text{Utility}_t^{b,2}}\xspace}

\newcommand{\T}{\ensuremath{T_u}\xspace}

\newcommand{\E}[1]{\mathbb{E}\left[#1\right]}
\newcommand{\prob}[1]{\mathbb{P}\left[ #1 \right]}

\newcommand{\uoptimal}{\ensuremath{U_{\text{opt}}}\xspace}

\newcommand{\optimalityName}{optimality index\xspace}
\newcommand{\OptimalityName}{Optimality Index\xspace}
\newcommand{\optimalityNotation}{\ensuremath{OI_{\protocol}}\xspace}

\newcommand{\adversarialSus}{adversarial susceptibility\xspace}
\newcommand{\AdversarialSus}{Adversarial Susceptibility\xspace}
\newcommand{\adversarialNot}{\ensuremath{AS_{\protocol}}\xspace}

\newcommand{\equilibriumTime}{\ensuremath{\mathcal{T}_e}\xspace}

\newcommand{\deltalmat}{\ensuremath{\mathbf{\Delta L}}\xspace}
\newcommand{\thetalvec}{\ensuremath{\mathbf{\Theta}_l}\xspace}
\newcommand{\thetahatlvec}{\ensuremath{\mathbf{\hat{\Theta}}_l}\xspace}
\newcommand{\plmat}{\ensuremath{\mathbf{P}_l }\xspace}
\newcommand{\epsilonvec}{\ensuremath{\bm{\varepsilon}}\xspace}

\newcommand{\deltabmat}{\ensuremath{\mathbf{\Delta B}}\xspace}
\newcommand{\thetabvec}{\ensuremath{\mathbf{\Theta}_b}\xspace}
\newcommand{\thetahatbvec}{\ensuremath{\mathbf{\hat{\Theta}}_b}\xspace}
\newcommand{\pbmatt}[1]{\ensuremath{\mathbf{P}_b^{#1}}\xspace}
\newcommand{\pbmat}{\ensuremath{\mathbf{P}_b}\xspace}

\newcommand{\Tmarket}{\ensuremath{T_{m}}\xspace}


%% file: introduction.tex
\section{Introduction}

\noindent\textbf{Lending in DeFi} Decentralized Finance (DeFi) has revolutionized lending and borrowing by eliminating centralized intermediaries. The main paradigm shift has been around moving away from opaque financial entities such as banks, that use proprietary models and data to match deposits with borrowers \cite{propModels}, to transparent pools with published algorithms to change interest rates, and borrowing conditions. Major DeFi lending platforms like Aave \cite{aaveWP} and Compound \cite{compoundWP} function through these liquidity pools, where lenders provide capital that borrowers can access. These protocols ensure that borrowers pledge enough collateral to cover their debt, along with an additional safety buffer.

\noindent\textbf{Stable utilization} The simplest variable of interest that any lending protocol seeks to control is the supply-demand ratio of the pool, referred to as ``utilization.'' The objective is to maintain a stable utilization around a designated ``optimal utilization'' threshold. A coarse rule of thumb is that when the utilization is low, interest rates remain low to encourage borrowing \cite{irThumbRule}. As utilization increases, interest rates rise to balance demand with supply and to prevent excessively high utilization, which could restrict lenders' ability to withdraw their funds, thus rendering the market less attractive.

\noindent\textbf{Collateral factor} Besides the interest rate, other parameters like the over-collateralization ratio, also known as the ``collateral factor,'' govern the long-term risks and profits of the market \cite{riskColDeFi}. In particular, the cash flow that any lender gets from the protocol is at risk of liquidation, and in the more severe cases, default. This risk can be minimized by demanding a large amount of collateral from borrowers, which makes the risk vanishingly small while making the lending market incredibly inefficient and unattractive for borrowers, especially if the asset used as the collateral does not suffer frequent price fluctuations. Thus, the collateral factor needs to be determined based on a careful analysis of recent historic behavior of the collateral asset price, and the risk appetite of the lender.

\noindent\textbf{Shortcomings of existing approaches} Present DeFi platforms fix the interest rate as a static function of the utilization \cite{aaveIR,compoundIR}. Choosing utilization as the primary indicator of both supply/demand dynamics and market risk/attractiveness and employing a fixed interest rate curve to manage these aspects is very arbitrary and manually determined. Furthermore, traditional DeFi borrowing and lending markets set the collateral factor through a comprehensive process involving community proposals and review phases \cite{aaveGov,compoundGov}. However, this method is notably slow and struggles to adapt quickly to rapid market changes, potentially leading to losses and excessive risks due to the delayed adjustment of parameters in response to market fluctuations and experiencing long periods of extreme low liquidity or market inefficiencies. For instance, the authors of \cite{gudgeon2020defi} have found that the markets for DAI and USDC frequently exhibit periods of extreme low liquidity with utilization exceeding 80\% and 90\%, respectively, which further highlights the inadequacy of current interest rate models.

\noindent\textbf{Need for adaptivity and fairness} In this study, we propose an adaptive, automated, and data-driven approach for designing a borrowing/lending protocol. We begin by modeling the behaviors of borrowers and lenders based on their incentives in a principled manner and examine how external factors alter these behaviors over time. We then define market equilibrium (Definition \prettyref{def:eq}), where the market remains stable within a broader external market, and rates offered by our protocol do not allow borrowers or lenders to gain an advantage over external market rates. Achieving equilibrium is crucial in a market with conflicting interests, as instability tends to disproportionately benefit one group over another, reducing overall fairness and attractiveness of the market \cite{marketDistortion}. If a protocol cannot dynamically adjust to achieve equilibrium, it risks losing liquidity and users. Market stability must be promptly restored after disruptions, which may be caused by changes in external market conditions or shifts in price distributions that affect the market's risk and profit structure.

\noindent\textbf{Thinking fast - convergence to equilibrium } Traditional stationary-curve borrowing/lending markets depend on user interactions to push towards equilibrium—for instance, high interest rates prompt borrowers to repay loans, reducing utilization and interest rates. However, this process is slow and often results in impermanent loss \cite{impLossLending}, especially for users with less flexibility in managing their assets. In order to address this problem, our protocol includes an ``interest rate controller'' submodule that learns the equilibrium interest rate from user behaviors, providing a faster convergence rate, even in the presence of uninformed users who lack precise information on competitive rates. Unlike traditional methods, our approach does not solely rely on user actions but actively learns from them to accurately assess and adapt to market conditions. Moreover, we provide an adversarial robust version of our interest rate controller as well which learns the equilibrium interest rate as long as the adversary controls less than 50\% of the borrowing demand.

\noindent\textbf{Thinking slow - adapting to the broader market} In addition to promptly restoring stability following market disruptions, it is essential for a protocol to adaptively optimize hyperparameters that enhance long-term system efficiency and manage risk. Our protocol features a long-term planner that uses the collateralization ratio as a control variable to adapt to market changes and stabilize the market at a desired level (\prettyref{sec:collateral_planner}). The collateralization ratio is critical for managing long-term risks and rewards in the system. This ratio has a complex relationship with user behavior and the overall risk and profitability of the market, which we explore in detail in our paper (\prettyref{sec:eval}). The objectives of the protocol within this long-term planner can be defined in many different ways; In this paper our focus is on maintaining long-term utilization at a target level and controlling default, however more complex objective functions could be implemented to address specific market needs or objectives. Our approach provides a general framework for designing adaptive markets with heterogeneous users who may have varying incentives.

\noindent\textbf{Evaluation} Additionally, we implemented our protocol and tested it with simulated borrowers and lenders, empirically comparing its performance against fixed-curve baselines. We evaluated the correctness of our theoretical guarantees in practice and demonstrated that our interest rate controller can quickly learn the equilibrium interest rate after each market disruption, regardless of borrowers' and lenders' elasticity. In contrast, the baseline protocol fails to find the equilibrium interest rate when user elasticity is low due to its reliance on user reactions to push the market toward equilibrium. Moreover, we showed that in the presence of major market changes, our protocol's collateral factor planner adaptively activates. By learning new price and market parameters, it sets the collateral factor to maintain utilization near a predefined optimal level in the long term.
\subsection{Related Work}

\noindent\textbf{Lending in DeFi } Various models on lender and borrower behavior and their equilibria have been explored. \cite{cohen23} assume parametrized supply and demand curves based on interest rates, approximating the curve around equilibrium to recommend rates, but they ignore external markets and default risk minimization. \cite{rivera2023equilibrium} consider external markets, measuring protocol efficiency by interest rate differences, but their models lack long-term decision-making and liquidation considerations. \cite{chiu2022fragility} examines Nash equilibrium in a model with independent quality shocks, showing that exogenous asset prices yield one equilibrium, while protocol-influenced prices cause oscillation and propose ad-hoc contract adjustments. Empirical studies on lender/borrower behavior \cite{aave_gauntlet,compound_gauntlet,compoundEmpirical} inform our parameter values. \cite{szpruch2024leveraged} discusses borrower trading strategies with market makers. Adversarial attacks on lending protocols have also been highlighted \cite{chitra2023attacks,cohen2023paradox,carre2023security}.

\noindent\textbf{Adaptive DeFi protocols} Several recent works in the mechanism design of financial systems have been advocating for the use of automated adaptivity \cite{AIeconomist}. The presence of impermanent loss and arbitrage loss in the design of market makers has also spawned multiple works in adaptive market making \cite{goyal2023finding, milionis2023automated, nadkarni2023zeroswap}. We seek to bring similar automated methods to lending in DeFi. Platforms such as Morpho \cite{morpho} and Ajna \cite{ajna} provide lenders and borrowers more flexibility when it comes to equilibrium interest rate discovery via an order-book like structure. However, such protocols require constant monitoring on the part of the participants for fairness and optimality. Our objective is to bring these notions of fairness/optimality to more passive pool-based lending protocols.

\noindent\textbf{Control} The methods used in this work are based on optimal control and filtering literature using the least squares method \cite{kailath1980linears}. This method has been used in the estimation of underlying dynamics, given a noisy access to measurements \cite{kuo1998least,zhang2020adaptive}. Several recent works have provided extensions of this algorithm to ensure adversarial robustness \cite{bhatia2015robust, xing2021adversarially, mcwilliams2014fast}, which use hard thresholding and concentration inequalities to weed out adversarial data. 

%% file: Problem_Formulation.tex
\section{Problem Formulation}
\subsection{Market Actors}

The DeFi borrowing and lending market includes four key participants: lenders, borrowers, liquidators, and the protocol, here called \protocol. These actors interact within a shared pool. This subsection briefly clarifies each participant's role and the mechanisms protocols use to regulate their interactions.

\noindent\textbf{Liquidators} To prevent defaults during price declines, the protocol employs \emph{liquidation}. This occurs when a borrower's \emph{loan-to-value} ratio (debt-to-collateral value) exceeds a threshold \liqthrsh (\(\liqthrshN\)), set between \(0\) and \(1\) and higher than the initial loan-to-value ratio \(c\). When this threshold is surpassed, liquidators can claim a portion of the borrower's collateral to repay the debt, reducing the loan-to-value ratio. Liquidators receive a fee \(\liqIncntive\) from the borrower's collateral. Liquidations enhance system safety but are unfavorable for borrowers due to the incentive fee. This prompts borrowers to increase their collateral preemptively. Despite liquidation mechanisms, defaults can occur if collateral prices drop abruptly or if liquidators lack sufficient incentives to act.

\noindent\textbf{Protocol} The protocol must adjust parameters \(\{r_t, \colfacN_t, \liqthrshN_t, \liqIncntive_t\}\) over time to stabilize the pool. Objectives include stabilizing loan supply and demand by setting an interest rate \(r_t\) and optimizing parameters to minimize defaults and liquidations while maintaining an ideal utilization rate. Our paper focuses on creating a competitive DeFi protocol with efficient rates, not on revenue maximization. The openness of DeFi protocols and minimal fees should ensure that the most competitive protocol eventually dominates the market.

\noindent\textbf{Asset pools }The lending pool consists of two assets: a stable asset, \assetOne, provided by lenders for interest, and a volatile asset, \assetTwo, used by borrowers as collateral. Borrowers can only borrow a fraction \colfac (\(\colfacN\)) of their collateral, set by the \paramProt. At timeslot \(t\), the overall pool's assets of type \assetOne, considering both lent-out funds and available liquidity, are denoted by \(\lt\). Note that \(\lt\) increases over time as lenders accrue interest on the lent-out portion. The asset of a particular lender \(i\) is represented by \(\lt(i)\).

\noindent\textbf{Borrowers and lenders }The interest rate \(\rt\) is set by the \paramProt at each block. Borrowers pay this rate, but lenders earn interest only on the utilized fraction \(\ut\) of their deposit, defined as \(\ut = \frac{\bt}{\lt}\), where \(\bt\) represents the overall debt across all borrowers, and \(\bt(i)\) represents the debt of borrower \(i\). The debt amount also increases over time due to accrued interest. The quantity of the overall collateral posted by all borrowers is denoted by \(\ct\), and \(\ct(i)\) denotes the collateral of borrower \(i\). Hence, \(\pt \cdot \ct\) determines the value of the collateral in terms of the lent-out asset, \assetOne (for a thorough list of the notations and their description refer to Appendix \ref{app:notation-table}). When borrowers repay, they return the loan plus interest and retrieve their collateral. Lenders receive interest based on the protocol rate and fund utilization. Defaults can affect the final interest rate for lenders. If collateral value drops below the debt, the protocol cannot compel repayment of the insolvent debt, resulting in a loss that impacts the lenders' interest rate.

\subsection{Environment model}

\paragraph*{Asset price model} We operate within discrete time intervals, denoted as \(\Deltat\), each corresponding to one blocktime. 
 We use a discrete price model to monitor the collateral asset's price from one block to the next. For simplicity, we assume the lent-out asset is a stablecoin with a relatively stable price, while the collateral asset's price follows an exogenous geometric Brownian motion with volatility \(\volat\). We assume constant volatility over short periods of time, with occasional sporadic jumps, but no fluctuations from one timeslot to the next. In particular, we assume that the price volatility is constant within timescales denoted by $\Tmarket > 1$ ($m$ for  market, denoting the timeframe within which the market is stable) which consists of multiple timeslots and can change arbitrarily every \Tmarket timeslots.

\noindent\textbf{Geometric Brownian price} The price at time \(t\), denoted as \(\pt\), follows a Geometric Brownian motion with drift \(\mu_{\text{price}}\) and volatility \(\volat\). The initial price \(p_0\) is the starting point, for notation simplicity we normalize and consider $\Deltat=1$ and hence formally, the model is:
\begin{equation}
p_t = p_{t-1} \exp\left(\muprice  + \volat \varepsilon_t \right), \quad \varepsilon_t \sim \mathcal{N}(0, 1)
\end{equation}\label{eq:price-model}
\noindent where \(\varepsilon_t\) is the innovation term for the volatility. 

\paragraph*{External market competition}
We assume the existence of an external competitor market which offers risk-free borrow rate $\robt$ and risk-free lend rate $\rolt$. These rates are constant during each market period $\Tmarket$ and can change arbitrarily between periods. This assumption accounts for the competition and the broader market within which our protocol operates. $\rolt$ and $\robt$ might represent the existence of complex alternatives rather than simple risk-free rates. In Appendix \ref{appendix:subsec:competition}, we discuss interpreting these parameters based on real-world strategies and competitors in the Defi ecosystem. Throughout the paper, we abstract these concepts into $\rolt$ and $\robt$.

\subsection{Protocol behaviour and pool logic}\label{subsec:protocol-behaviour}

In this section, we establish the structure of a decentralized borrowing-lending protocol, denoted by \protocol. The protocol fulfills two primary roles: 1) \protocol sets the pool's parameters for each block, denoted as \(\{r_t, \colfacN_t, \liqthrshN_t, \liqIncntive_t\}\), by transmitting a transaction to the underlying blockchain. These parameters govern the pool's logic. 2) \protocol updates the state variables \(\lt\), \(\bt\), and \(\ct\) every block to apply interest rate accumulation and liquidation or default due to price fluctuations.

\noindent\textbf{Handling default} At the beginning of each timeslot \(t\), \protocol receives the latest price of \assetTwo, \(\pt\), from an oracle. The protocol calculates potential defaults accrued in the last timeslot for each user. The default for borrower \(i\) at timeslot \([t-1,t]\) is:
\begin{equation}\label{eq:userDefault}
    \userDefault{t-1}{i}(\pt) \coloneqq \max\left\{0, \bS{t-1}(i) - \cS{t-1}(i) \cdot \pS{t}\right\}
\end{equation}

\noindent The overall default, normalized by \(\lS{t-1}\), is:
\begin{equation}\label{eq:poolDefault}
     \poolDefault{t-1}(\pt) \coloneqq \frac{1}{\lS{t-1}}\sum_{i \in \mathrm{borrowers}} \max\left\{0, \bS{t-1}(i)-\cS{t-1}(i)\cdot \pS{t}\right\}
\end{equation}
 The protocol seizes the remaining collateral of defaulted positions, exchanges it for \assetOne, and sets the debt of defaulted borrowers to zero. The gained \assetOne assets are added back to the pool.
 Moreover the underwater debt is deduced from the lender's deposit accordingly:
\begin{equation}
    \lt(i) = \lS{t-1}(i) - \poolDefault{t-1}(\pt) \cdot \lS{t-1}(i), \quad \forall\, i \in \mathrm{Lenders}
\end{equation}
For a more thorough explanation of why we handle defaults in this way rather than using a safety reserve similar to most of the current working Defi borrow-lending platforms refer to Appendix \ref{app:protocol-logic}.

\noindent\textbf{Interest update} The debt of non-defaulted borrowers is updated by:
\begin{equation*}
    \bt(i) = \bS{t-1}(i) \cdot \left(1 + \rS{t-1} \right), \quad \forall \,i \in \mathrm{Borrowers}
\end{equation*}

\noindent The interest rate on the utilized portion of the pool applies to the lenders as well:
\begin{equation*}
    \lt(i) = \lt(i) \cdot \left(1 + \rS{t-1}\, \frac{\bS{t-1}}{\lS{t-1}} \right), \quad \forall \,i \in \mathrm{Lenders}
\end{equation*}

\noindent\textbf{Liquidation} \protocol tracks borrow positions exceeding the liquidation threshold. A position \(i\) is eligible for liquidation if \(\liqthrshN_{t-1} < \frac{\bt(i)}{\ct(i) \cdot \pt} < 1\). Liquidators reduce the user's debt by purchasing collateral, restoring the loan-to-value ratio below \(\liqthrshN_{t-1}\). The debt and collateral after liquidation are updated accordingly. The minimum liquidation amount that reduces the user's loan-to-value below $\liqthrshN_{t-1}$ is

\begin{equation*}\label{eq:userLiq}
    \userLiq{t-1}{i}(\pt) \coloneqq \max\left\{0, \frac{\bt(i)-\liqthrshN_{t-1} \cdot \ct(i) \cdot \pt}{1-\liqthrshN_{t-1}(1+\liqIncntive_{t-1})}\right\}
\end{equation*}

\noindent\textbf{Setting new parameters} \protocol sets new parameters for the next timeslot: \(\{r_t, \colfacN_t, \liqthrshN_t, \liqIncntive_t\}\). These parameters determine the interest rate, maximum loan-to-value, and liquidation parameters for the next timeslot.

\noindent\textbf{Admitting new users} The protocol accepts all new lend and repay requests. Borrow requests are accepted only if they adhere to the maximum collateralization factor $\colfacN_t$ and there is sufficient \assetOne in the pool to satisfy the request. Additionally, it processes withdrawal requests from lenders as long as the withdrawal amount does not exceed the available \assetOne in the pool.


\subsection{User behavior model}\label{subsec:user-behaviour-model}

In this section, we establish the model that a rational user would use to interact with the protocol.

\noindent\textbf{Continuum of users} We consider a continuum of lenders and borrowers, each controlling a single unit of demand or supply. In each timeslot, lenders can deposit or withdraw their unit, and borrowers can borrow, repay, or adjust their loan-to-value ratio by sending a transaction to the smart contract.



\subsubsection{Lender}

\noindent\textbf{Utility function} We assume that a continuum lender with one unit of supply, planning for the next timeslot, will calculate the following utility function at time $t$ to decide whether to deposit into the pool or, if already deposited, to withdraw and invest in another external alternative offering \(\rolt\).
\begin{equation}\label{eq:lender-effective-r}
    \utilityl \coloneqq \rt\ut - \E{\poolDefault{t}(\pS{t+1})} - \rolt
\end{equation}
\noindent where the expectation is over the price at timeslot $t+1$ price. The first term in \ref{eq:lender-effective-r} represents the interest earned by the lender on the utilized portion of their deposit. Although the interest rate is compounded, we approximate it linearly for simplicity. 
The second term denotes the normalized defaulted debt deducted from the lender's deposit. 
 Finally, we subtract the external interest rate \(\rolt\) to account for the lender's opportunity cost.



\noindent\textbf{Lending dynamics} We now describe the dynamics by which lenders add or withdraw their deposits based on utility. We introduce a new parameter \(\elasticityl\), which reflects the average elasticity of lenders at time \( t \). This value can change over time (not faster than \Tmarket) and is unknown to the protocol. We assume that the relative rate at which lenders deposit or withdraw from the pool is governed by the following model:
\begin{align}\label{eq:dynamic-lender}
    \frac{\lS{t+1} - \lt}{\lt} &= \elasticityl \cdot \utilityl + \varepsilon_t \notag \\ 
    &= \elasticityl \cdot \left( \rt\ut - \mathbb{E}[\poolDefault{t}(\pS{t+1})] - \rolt \right) + \varepsilon_t, \qquad \varepsilon_t \overset{\text{i.i.d}}{\sim} \mathcal{N}(0, \zeta^2)
\end{align}

The noise term accounts for the behavior of uninformed or less informed users.

\subsubsection{Borrower}

From observing the real lending markets, we identify two types of DeFi borrowers, financing and leveraged trading borrowers (see Appendix \ref{appendix:subsec:borrower-behaviour} for more details).

\noindent\textbf{Financing borrowers} The first type borrows an asset to use elsewhere, gaining value by leveraging it, e.g., for yield farming or real-world purposes. This group's value from the borrowed asset is represented as \(\robt\), measured as an interest rate. The utility function of a borrower of this type controlling one unit of demand, considering the opportunity cost of locked collateral, is
\begin{align}\label{eq:borrower-one-effective-r}
    \utilitybOne \coloneqq \robt - \rt + \mathbb{E}\left[\userDefault{t}{i}(\pS{t+1})\right] 
    &- \mathbb{E}\left[\userLiq{t}{i}(\pS{t+1})\right] \cdot \liqIncntive_t \nonumber \\ 
    &+ \ct(i) \cdot \mathbb{E}\left[\indicator{\pS{t+1} - \pt < 0}(\pS{t+1} - \pt)\right]
\end{align}

\noindent This includes inherent value (\(\robt\)), interest rate (\(-\rt\)), default value (\(\E{\userDefault{t}{i}(\pS{t+1})}\)), liquidation cost (\(-\E{\userLiq{t}{i}(\pS{t+1})} \cdot \liqIncntive_t\)), and opportunity cost of locked collateral $\mathbb{E}\left[\indicator{\pS{t+1} - \pt < 0}(\pS{t+1} - \pt)\right]$.

\noindent\textbf{Leveraged trading borrowers} The second type aims to take a long position on \(\assetTwo\). They borrow \(\frac{1}{\pt}\) units of \(\assetTwo\) from an external provider \(\mathcal{Z}\), add more collateral to meet the over-collateralization requirement $\colfacN_t$ , and borrow one unit of \(\assetOne\) from \protocol, then exchange it for \(\assetTwo\) to repay \(\mathcal{Z}\). This borrowing strategy is studies in details in \cite{szpruch2024leveraged}. Their utility function is:
\begin{equation}\label{eq:borrower-two-effective-r}
    \utilitybTwo = -\rt + \E{\userDefault{t}{i}(\pS{t+1})} - \E{\userLiq{t}{i}(\pS{t+1})} \cdot \liqIncntive_t + \E{\frac{\pS{t+1} - \pt}{\pt}}
\end{equation}
\noindent This includes interest rate (\(-\rt\)), default value (\(\E{\userDefault{t}{i}(\pS{t+1})}\)), liquidation cost (\(-\E{\userLiq{t}{i}(\pS{t+1})} \cdot \liqIncntive_t\)), and gain from a price change of the $\frac{1}{\pt}$ investment in \assetTwo which was possible through interacting with \protocol i.e.,(\(\E{\frac{\pS{t+1} - \pt}{\pt}}\). Refer to Appendix \ref{appendix:subsec:borrower-behaviour} for more details.

\noindent\textbf{Borrowing dynamics} Assuming \(\alpha\) fraction of borrowers are type 1 and the rest type 2, the rate of borrowing or repaying is a linear function of borrower's elasticity \(\elasticityb\) and their average utility plus noise:
\begin{align}\label{eq:dynamics-borrower}
    \frac{\bS{t+1} - \bt}{\bt} &= \elasticityb \cdot \big(\alpha\cdot\utilitybOne + (1-\alpha)\cdot \utilitybTwo \big) + \varepsilon_t \notag
    \\&= \elasticityb \cdot \bigg( \alpha\cdot\robt-\rt + \E{\userDefault{t}{i}(\pS{t+1})} - \E{\userLiq{t}{i}(\pS{t+1})}\cdot \liqIncntive_t\notag\\&+ \alpha\cdot  \ct(i) \E{\indicator{\pS{t+1}-\pt<0}(\pS{t+1}-\pt)} + (1-\alpha) \cdot\E{\frac{\pS{t+1} - \pt}{\pt}} \bigg)+\varepsilon_t,
\end{align}
where $\varepsilon_t\overset{\text{i.i.d}}{\sim}  \mathcal{N}(0,\zeta^2)$.

\noindent\textbf{Collateralization strategy}
Throughout this paper, we assume that \protocol\ efficiently sets the buffer between the collateral factor and the liquidation threshold to ensure that the expected liquidation, \(\mathbb{E}[\userLiq{t}{i}(p_{t+1})]\), for a borrower $i$ who maintains the posted collateral factor, is negligible. This assumption differs from the current borrowing and lending platforms, which experience significant liquidations even among borrowers who adhere to the posted collateral factor. However, we believe that a competent borrowing and lending protocol should set risk parameters to minimize this risk.
In Section~\ref{sec:collateral_planner}, we explain how we determine the liquidation threshold and collateral factor to ensure that the expected liquidations remain negligible.
Moreover, throughout this paper, we assume that the liquidation incentive, \(\liqIncntive_t\), is set sufficiently high by the protocol to encourage the borrowers to maintain the collateral factor, \(\colfacN_t\), and avoid liquidations.

\begin{lemma}[Maximum loan-to-value adoption]\label{lemma:collateral-strategy}
    Consider the following conditions:
    \begin{itemize}
        \item The collateral factor, $\colfacN_t$, and the liquidation threshold, $\liqthrshN_t$, are chosen such that for a given $\liqthrshN_t$, $\colfacN_t$ is the maximum collateral factor that ensures the expected liquidation, $\mathbb{E}[\userLiq{t}{i}]$, is approximately zero for a user $i$ who maintains $\colfacN_t$.
        \item The liquidation incentive, $\liqIncntive_t$, is set high enough to incentivize rational borrowers to avoid liquidation by ensuring that $\frac{\bt(i)}{\ct(i)\pt} \leq \colfacN_t$.
    \end{itemize}
    Then rational borrowers will adopt the maximum loan to value allowed by the protocol i.e., $\colfacN_t$.
\end{lemma}

Hence from now on, we assume that $\mathbb{E}[\userLiq{t}{i}]$ for a rational continuum borrower is negligible and for ease of notation, we denote it by $\userLiq{}{}(\colfacN_t,\liqthrshN_t)$.

\subsubsection{Liquidator}
We assume that the liquidation incentive 
\liqIncntive is set at a level that consistently incentivizes liquidators, ensuring their prompt engagement and immediate liquidation up to the limit allowed by \protocol. Additionally, since we use an exogenous price model for collateral, phenomena like liquidation spirals studied in previous works \cite{klages2021stability} are not considered in our analysis. 

\subsection{Equilibrium analysis}

In this section, we will formally define the concept of equilibrium in the borrow-lending framework we established. And we will analytically identify the set of equilibria of this market when users follow the behaviour outlined in \ref{subsec:user-behaviour-model}.


\begin{definition}[Market equilibrium]\label{def:eq}
A lending pool governed by protocol \protocol parameterized by \(\{\rt, \colfacN_t, \liqthrshN_t, \liqIncntive_t\}\), and lender's and borrower's behavior respectively governed by \ref{eq:dynamic-lender}, and \ref{eq:dynamics-borrower} 
is in equilibrium if and only if:
\begin{equation*}\label{eq:equilibrium-condition}
\E{ \frac{\bS{t+1} - \bt}{ \bt} } =0 \quad \text{and} \quad \E{\frac{\lS{t+1} - \lt}{\lt} } =0
\end{equation*}
where the expectation is over the noise term in the user behavior model.
\end{definition}

\begin{lemma}[Simplified default and price change terms ]\label{lemma:simplify-default-to-the-end}
    In the presence of rational continuum lenders and rational continuum borrowers who follow collateral factor $\colfacN_t$ (due to Lemma \ref{lemma:collateral-strategy}) we have:
         \begin{align}\label{eq:simplify-default-to-the-end}
        \poolDefault{}{}(\colfacN_t) &\coloneqq \E{\userDefault{t}{i}(\pS{t+1})} \notag\\&= \Phi\left(\frac{\log(\colfacN_t) - \mu}{\sigma}\right) - \frac{\exp{(\frac{\sigma^2}{2}+\mu)}}{\colfacN_t}\cdot\Phi\left(\frac{-\mu+\log(\colfacN_t)-\sigma^2}{\sigma}\right)
    \end{align}
    \begin{equation}
        \E{\poolDefault{t}(\pS{t+1})} = \ut \poolDefault{}(\colfacN_t)
    \end{equation}
    \begin{equation}
        \ct(i) \E{\indicator{\pS{t+\T} - \pt < 0} (\pS{t+\T} - \pt)} = \frac{1}{\colfacN_t} \left(e^{\mu + \frac{\sigma^2}{2}} \frac{\Phi\left(\frac{-\muprice - \sigma^2}{\sigma}\right)}{\Phi\left(\frac{-\muprice}{\sigma}\right)} - 1\right)
    \end{equation}
    \begin{equation*}
        \E{\frac{\pS{t+1} - \pt}{\pt}} = e^{\muprice + \frac{\sigma^2}{2}} - 1
    \end{equation*}
    \noindent Where \(\phi(.)\) and \(\Phi(.)\) denote the PDF and CDF, respectively, of the standard normal distribution.
\end{lemma}
Throughout the rest of the paper, we denote the expected default for one unit of debt by \(\poolDefault{}{}(\colfacN_t)\). Conceptually, Lemma \ref{lemma:simplify-default-to-the-end} implies that if the pool's loan-to-value ratio is maintained close to \(\colfacN_t\) in each timeslot, the probability of default remains memoryless. This is due to the Brownian motion of price, where the price ratio between consecutive timeslots follows a stationary distribution. Therefore, the probability that the next timeslot's price falls below the default threshold is stationary and memoryless, changing only when the price distribution itself changes.

\begin{theorem}\label{theorem:equilibrium-finding}
Let the lender behavior be described by Equation \ref{eq:dynamic-lender}, the borrower behavior by Equation \ref{eq:dynamics-borrower} (for $\elasticityb > 0$), and the assumptions of Lemma \ref{lemma:collateral-strategy} hold. A protocol with parameters \(\{\rt, \colfacN_t, \liqIncntive_t, \liqthrshN_t\}\), achieves a non-trivial equilibrium if and only if \(\rt=\rS{}^*\):
\begin{align}\label{eq:equilibrium-rate}
\rS{}^* &= 
\alpha \,\robt + \poolDefault{}(\colfacN_t) + \frac{\alpha}{\colfacN_t} \left(e^{\mu + \frac{\sigma^2}{2}} \frac{\Phi\left(\frac{-\mu - \sigma^2}{\sigma}\right)}{\Phi\left(\frac{-\mu}{\sigma}\right)} - 1\right) + (1 - \alpha) \left(e^{\mu + \frac{\sigma^2}{2}} - 1\right)
\end{align}

\noindent Furthermore, if $\elasticityl>0$, then the unique equilibrium utilization is:

\begin{equation}\label{eq:equilibrium-u}
\uS{}^* =
\begin{cases}
    \frac{\rolt}{\rS{}^* - \poolDefault{}(\colfacN_t)} & \text{if }\rS{}^* > \poolDefault{}(\colfacN_t) \\
    1 & \text{otherwise}
\end{cases}
\end{equation}
\end{theorem}

\noindent\textbf{Equilibrium dynamics} In order to achieve the equilibrium point, The protocol first should find the equilibrium interest rate, \(r^*\), and set $\rt = r^*$ to prevent borrowers from leaving or joining the system. Once \(r^*\) is set, if the utilization \(U_t\) is above \(U^*\), lenders' utility is positive, so they keep lending until \(U_t\) reaches \(U^*\), stabilizing the system. Conversely, if \(U_t\) is below \(U^*\), lenders' utility is negative, causing them to withdraw until \(U_t = U^*\). At this point, lenders are indifferent between the pool and external competitors and remain fixed. The equilibrium utilization is \(U^* = 1\) if, regardless of utilization, lenders' utility under \(\rt = r^*\) remains non-positive. Consequently, they withdraw fully, yielding \(U^* = 1\), and they remain trapped in an unfavorable equilibrium where they prefer external rates but since their fund is being borrowed, they cannot leave the system.

\subsection{Protocol objectives and evaluation metrics}

In this section, we define three key metrics to evaluate a borrow-lending protocol. The protocols considered follow the behavior outlined in \ref{subsec:protocol-behaviour}. Each protocol selects a deterministic or randomized interest rate function \(\rt: \mathcal{H}_1^t \to \mathbb{R}^+\), a collateral factor function \(\colfacN_t: \mathcal{H}_1^t \to [0, 1]\) and a liquidation threshold function \(\liqIncntive_t: \mathcal{H}_1^t \to [0, 1]\) mapping the pool's history to an interest rate, a collateral factor and a liquidation threshold.

\subsubsection*{Rate of equilibrium convergence}

\noindent\textbf{The importance of stability} Achieving market equilibrium is crucial for any DeFi application, as equilibrium points represent the market's most competitive state. At equilibrium, no participant is overpaid or underpaid, ensuring all users are equally satisfied with \protocol{} as with any external alternative.
In two-sided markets like borrow-lending, the time to achieve stability can disproportionately benefit one side. Setting interest rates below what borrowers are willing to pay results in lenders receiving less than in a competitive market. Conversely, if interest rates exceed the equilibrium rate, borrowers pay more than in a stable market. The more elastic user benefits at the expense of the less elastic user, who incurs an impermanent loss.

\noindent\textbf{Adapting to market changes} Adapting to market changes allows the protocol to stabilize the pool when user behavior or price volatility changes. As these parameters change over time, the protocol must respond dynamically to maintain system stability within a reasonable timeframe.
In the borrow-lending market framework, each time user behavior parameters (e.g., \(\robt, \rolt, \alpha\)) or price volatility (\(\volat\)) change, the market stabilizes at a new \(\rS{}^*\) (refer to Theorem \ref{theorem:equilibrium-finding}). One objective of \protocol is to rapidly identify and set this new equilibrium rate. We formalize this concept as the rate of convergence, using it as a metric to evaluate our protocol against non-learning baselines.

\begin{definition}[Rate of equilibrium convergence] 

\noindent Consider a stable borrow-lending pool with the interest rate \(\rt = \rS{}^*\), where \(\rS{}^*\) is the equilibrium interest rate determined by Equation \ref{eq:equilibrium-rate}. At time \(t+1\), user behavior models adapt to new parameters \(\{\bar{\elasticityl}, \rolbar, \bar{\elasticityb}, \robbar, \alpha\}\), and price volatility changes to \(\bar{\volat}\). Let \(\rS{\tau}\) represent the interest rate set by \protocol at time \(\tau > t\), and let \(\bar{r}^*\) denote the new equilibrium interest rate for the updated market parameters. We define the \emph{rate of equilibrium convergence} of \protocol, denoted by \(\convrate\), as the infimum of functions \(f(\delta, \tau)\), such that there exists some \(\mathtt{T}(\delta)\) for which, with probability \(1 - \delta\),
\[
|\rS{\tau} - \bar{r}^*| < f(\delta, \tau),\quad \forall \,\tau > \mathtt{T}(\delta),
\]
\noindent for any initial and secondary set of parameters in the user model and price model. Probabilities are calculated on the randomness of the protocol and the noise in the user behavior model.

\end{definition}

\subsubsection*{Equilibrium \OptimalityName}
\noindent\textbf{Protocol long-term objectives} Any Defi borrow-lending market aims to meet specific long-term objectives. For instance, the pool should maintain utilization at an optimal level; Because low utilization reduces capital efficiency, requiring the protocol to pay higher interest rates to lenders, and high utilization can make it difficult for lenders to withdraw and borrowers to secure loans. Moreover, expected defaults and liquidations are risk metrics that the protocol aims to control.
These objectives operate on a different timescale than the interest rate adjustments discussed in the convergence rate. For instance, we care about the average utilization or expected default over a long period rather than their local values in each timeslot.
To address this, we define a metric called \optimalityName, which evaluates the desirability of the system's equilibriums during the timeslots when the pool has reached its equilibrium, denoting the set of these timeslots by \equilibriumTime.

\begin{definition}[\OptimalityName]\label{def:optimality-index}

\noindent The \OptimalityName of a protocol, denoted by \optimalityNotation, is defined as follows:
\begin{equation}\label{eq:optimality-index}
    \optimalityNotation \coloneqq  \frac{1}{|\equilibriumTime|}\sum_{t \in \equilibriumTime}\E{-(\ut - \uoptimal)^2 - \gamma \,\big(\ut\,\poolDefault{}(\colfacN_t) + \userLiq{}{}(\colfacN_t,\liqthrshN_t)\big)},
\end{equation}

The expectation is taken over the protocol's randomness and user behavior. \uoptimal and \(\lambda\) are constant parameters.
\end{definition}

While the first term in the \optimalityNotation ensures that utilization is kept near a desired point, the second term acts as a regularization term, controlling the expected default and liquidation.
Our definition of \OptimalityName evaluates a specific notion of optimality, but it is just one of many possible definitions. Our protocol design methodology can be applied to other objective functions beyond Function \ref{eq:optimality-index}. In this paper, we showcase our ideas for this specific objective function and discuss how to extend the methodology to other objectives.

\subsubsection*{Adversarial robustness}

Protocols in DeFi are always susceptible to adversarial behavior that can be used to manipulate an adaptive algorithm to respond in a suboptimal manner. This behavior is usually observed when some borrower/lender agents interact with protocols outside of \protocol in conjunction with \protocol to achieve a profit. This can involve oracle manipulations attacks used to run away with valuable assets while providing worthless collateral, or attacks that move interest rates in the opposite direction of the equilibrium rates. We focus on the latter type of adversarial behavior in this work. Moving interest rates away from equilibrium leads to market inefficiencies. Further, undue hikes in these rates can be used to trigger unexpected liquidations for profit. 

We thus propose that the susceptibility of \protocol to adversarial manipulation should also be measured. To do that, we first assume that a fraction $\beta$ of the population of lenders/borrowers are adversarial. Thus, \protocol will face an adversarial lender/borrower for an approximately $\beta$ fraction of time slots. In those time slots, the adversary can manipulate the borrow/lend reserves arbitrarily. Let $T_\beta$ denote the set of time slots that the protocol faces an adversary.

We measure how susceptible a protocol is to adversarial manipulation using the following metric.

\begin{definition}[\AdversarialSus]\label{def:adversaria_sus}

The \adversarialSus of a protocol, denoted by \adversarialNot, is defined as follows:
\begin{equation*}
    \adversarialNot \coloneqq \E{\sum_{t \in \equilibriumTime} r_\protocol^t - r_{\protocol|T_\beta}^t}.
\end{equation*}
The expectation is taken over the protocol's randomness, user/adversarial behavior, where $r_\protocol^t$ denotes the interest rate recommended by the protocol and $r_{\protocol|T_\beta}^t$ is the same, when the indices of adversarial actions are known.
\end{definition}
Since an adversary can manipulate the protocol arbitrarily, the above measure signifies how adept the protocol is in weeding out historical data that has been manipulated, thus ensuring a cleaner convergence to the true interest rate equilibrium.
\subsection{Baseline}

For the baseline, we consider protocols akin to Compound, which utilize a piecewise linear interest rate curve to ensure stability. These protocols dynamically adjust interest rates at each block according to the model:
\begin{equation*}
R_t = 
\begin{cases} 
R_0 + \frac{U_t}{\uoptimal} R_{\text{slope1}}, & \text{if } U_t \leq \uoptimal \\
R_0 + R_{\text{slope1}} + \left(\frac{U_t - \uoptimal}{1 - \uoptimal}\right) R_{\text{slope2}}, & \text{if } U_t > \uoptimal
\end{cases}
\end{equation*}

\noindent In contrast to our proposed approach, these platforms generally set collateral factors and other market parameters through offline simulations that attempt to forecast near-future market conditions. Parameters are selected based on simulation outcomes and are subject to decentralized governance voting. Since this phase happens in an offline and opaque manner by centralized companies, we cannot compare this aspect of their protocol with ours.

\noindent\textbf{Limited user elasticity assumption} Due to the piecewise linear nature of the interest rate as a function of utilization, these protocols achieve market stability only when either borrowers or lenders, but not both, exhibit elasticity \cite{aave-report-2023}. To see why this is the case, note that according to Theorem \ref{theorem:equilibrium-finding}, if borrowers are elastic ($\elasticityb > 0$), the equilibrium interest rate can be uniquely determined. And if lenders are elastic too, the equilibrium utilization is determined by Equation \ref{eq:equilibrium-u} which is not a linear function of $r^*$, hence a piecewise linear interest rate curve cannot satisfy the equilibrium conditions if both sides are elastic. Traditional DeFi platforms typically monitor the pool to identify the more elastic side of the market (usually borrowers) and design the curve accordingly.

%% file: protocol-theoretical.tex
\section{Fast-slow thinker protocol}\label{sec:protocol-design-guarantees}

In this section, we outline a design for the interest rate and collateral factor function of \protocol that aims to achieve both the best possible convergence rate and the optimal \optimalityName. The protocol has the following two components.

\noindent \textbf{Interest rate online controller} A least squares estimator detects market disruptions that lead to instability and learns the equilibrium interest rate from borrowers' reactions, setting it agilely.

\noindent \textbf{Risk parameters planner} The long-term parameter planner consists of three parts:
1) A user behavior parameter estimator, which estimates the user behaviour model parameters that are required to optimize the \optimalityName.
2) An optimization module, which selects the optimal collateral factor to maximize \optimalityName, assuming negligible expected liquidation.
3) A liquidation threshold determination module, which sets the liquidation threshold as a function of the collateral factor to ensure zero expected liquidation.

\begin{figure}[htb]
  \centering
  \includegraphics[scale=0.7]{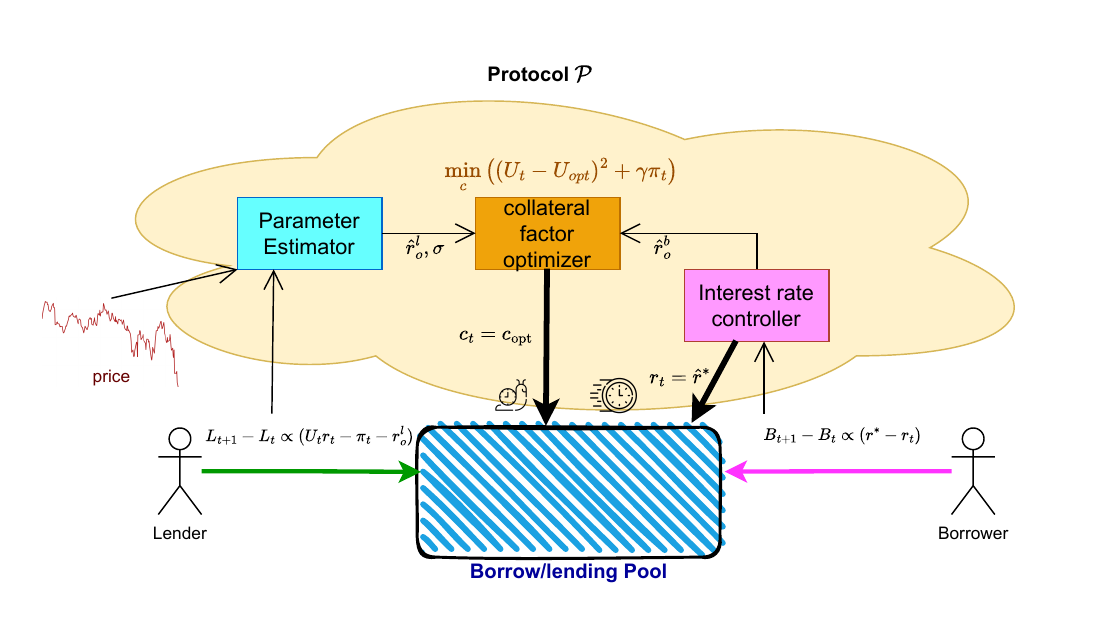}
  \caption{Protocol Overview. The interest rate controller observes borrower actions to estimate \(r^*\) and set \(\rt = \hat{r}^*\). The collateral factor planner includes a parameter estimator and an optimizer: the estimator finds \(\rolt\), \(\robt\), and \(\sigma\), while the optimizer uses these estimates to determine the optimal collateral factor for the market.}
  \label{fig:protocol-overview}
\end{figure}

The canonical scenario we use to evaluate our protocol is the following: at the beginning of the timeslot $t$, one or some of the market parameters (e.g., \(\volat, \rolt, \robt, \alpha\)) change to a new level and remain constant for a period \(\Tmarket\). The protocol must adapt to these new parameters by setting the equilibrium interest rate and the optimal collateral factor that maximizes the \optimalityName when the pool reaches equilibrium. Refer to Figure \ref{fig:protocol-overview} for a visual representation of the protocol and its interaction with the pool.


\subsection{Online interest rate controller}

\noindent\textbf{Linear regression} We model the problem of finding $r^*$ using the linear regression method, where the borrow/repay rate depends on the difference \(\rt - r^*\) (motivating factor for borrowers) and \(\elasticityb\) (borrower elasticity), with an added noise component; And use this model to estimate $r^*$ adaptively. Here is the linear regression problem formulation:
\begin{align*}
    \frac{\bS{t+1} - \bt}{\bt} &= \elasticityb \cdot \bigg( \alpha\,\robt-\,\rt + \poolDefault{}(\colfacN_t) + \frac{\alpha}{\colfacN_t}\, \E{\indicator{\pS{t+1}-\pt<0}(\frac{\pS{t+1}-\pt}{\pt})} + \\&(1-\alpha) \cdot\E{\frac{\pS{t+1} - \pt}{\pt}} \bigg)  \\&\overset{\ref{eq:equilibrium-rate}}{=}
    \elasticityb(\rS{}^* - \rt )
\end{align*}
\begin{equation}\label{eq:borrowers-ls-formulation}
\deltabmat = \pbmatt{t} \cdot \thetabvec + \epsilonvec
\end{equation}
where:
\begin{itemize}
    \item \(\deltabmat = \left[\frac{\bS{1} - \bS{0}}{ \bS{0}}, \frac{\bS{2} - \bS{1}}{\bS{1}}, \ldots, \frac{\bS{t+1} - \bS{t}}{ \bS{t}}\right]^T\) is the vector of normalized changes.
    \item $\pbmatt{t}$ is the matrix of pool variables that affect borrower's behaviour:
\[
(\pbmatt{t})^\top = \begin{bmatrix}
    1 & 1 & \cdots & 1 \\
    \rS{0} & \rS{1} & \cdots & \rt
\end{bmatrix}
\]
    \item \(\thetabvec = [\elasticityb\,\rS{}^*\;,-\elasticityb]^T\) is the parameter vector 
    \item \(\epsilonvec = [\varepsilon_{0}, \varepsilon_{1}, \ldots, \varepsilon_{t}]^T\) is the noise vector.
\end{itemize}

\noindent The interest rate controller algorithm, outlined in \ref{alg:interest-rate-controller}, activates when \(\Delta B_t\) exceeds a threshold \(\delta\), indicating \(\rt \neq r^*\). The algorithm collects \(\deltabmat\) and \(\pbmat^{t}\) to estimate \(\thetahatbvec\), setting \(\rt\) as the estimated \(\hat{r}^*\) according to \(\hat{r}^* = -\frac{\thetahatbvec(0)}{\thetahatbvec(1)}\).

\noindent\textbf{Exploration noise} We cannot use vanilla LSE in this setting and simply output \(\rt = \hat{r}^* = -\frac{\thetahatbvec(0)}{\thetahatbvec(1)}\) because feeding back the estimated \(\hat{r}^*\) to the protocol may cause the rows of the \(\pbmat{}\) matrix to become very close to each other. This happens when the estimator outputs an \(\rt\) that has been seen before, leading to redundant data points. Consequently, the theoretical guarantee of LSE to converge to the correct \(\thetabvec^*\) as the number of data points increases is impaired. To mitigate this problem, with a small probability, we sample a random interest rate. Finally, after running for a few iterations and accumulating a sufficiently large number of samples, the estimator stops and outputs the estimated \(r^*\).

\subsubsection{Theoretical analysis}
\begin{theorem}[LSE convergence rate]\label{theorem:lse-convergence-rate}
Assuming $\elasticityb> 0$, the interest rate controller described in Algorithm \ref{alg:interest-rate-controller} with stopping time $\tau$ (taking $\tau$ samples), satisfies the following:
\[
\convrate \sim \mathcal{O}\left(\frac{\log{\frac{1}{\delta}}}{\sqrt{\tau}}\right).
\]
Moreover, if $\elasticityb$ is known:
\(\E{\rS{\tau}} = r^*,\;\forall \tau\)
\end{theorem}

\begin{theorem}[Baseline convergence rate]
\label{theorem:baseline-conv-rate}
Under a piece-wise linear interest rate function 
\begin{itemize}
    \item In the presence of elastic borrowers and inelastic lenders, we have:
\begin{align*}
    &\E{\rS{\tau}} = r^* + D(1-K\,\elasticityb )^\tau ,\qquad 0<1-K\,\elasticityb<1\\ &\convrate \sim o\left(\frac{1}{\sqrt{\delta}}\right).
\end{align*}
\noindent where $K,D$ are constants determined from the specifications of the interest rate curve.
\item If lenders are elastic as well, the protocol never stabilizes with rate $r^*$.
\end{itemize} 
\end{theorem}

Our interest rate controller provides an unbiased estimation of the equilibrium interest rate, with the estimation variance decreasing over time. In contrast, the baseline algorithm is a biased estimator of the equilibrium rate, becoming unbiased only as \(\tau \to \infty\); And the rate of convergence of the average error to zero is proportional to \(\elasticityb\), as the baseline protocol relies heavily on user actions to adjust the rate toward equilibrium rather than actively learning from user behavior. 
Additionally, the baseline algorithm maintains a constant error relative to the equilibrium rate due to the noise in user behavior and its inability to filter out this noise. 

\input{algorithms/bestimator}

\subsection{Risk parameters planner}\label{sec:collateral_planner}
\noindent\textbf{Determining the liquidation threshold} First, we examine the conditions that ensure zero expected liquidation. These conditions provide the planner with the necessary bounds for setting the liquidation threshold.

 \begin{lemma}\label{lemma:liq-expectation}
The expected liquidation incurred at time \( t + 1 \), given that the loan-to-value ratio at time \( t \) is \(\colfacN_t\), can be expressed as 

\begin{align*}
\userLiq{}{}(\colfacN_t,\liqthrshN_t) &\coloneqq  E[\userLiq{t}{i}(\pS{t+1})] \\&= \frac{1}{1 - \liqthrshN_t} \left( \Phi \left( \frac{\ln \left( \frac{\colfacN_t}{\liqthrshN_t} \right) - \mu + \sigma^2}{\sigma} \right) - \frac{\liqthrshN_t}{\colfacN_t} e^{\mu + \sigma^2} \Phi \left( \frac{\ln \left( \frac{\colfacN_t}{\liqthrshN_t} \right) - \mu - \sigma^2}{\sigma} \right) \right)
\end{align*}
where \( \Phi \) denotes the cumulative distribution function of the standard normal distribution.

\end{lemma}

To maintain the expected liquidation below a small threshold (nearly zero), this lemma provides bounds on $\liqthrshN_t$ and $\frac{\colfacN_t}{\liqthrshN_t}$. These bounds are used by the planner to set $\liqthrshN_t$ and constrain $\colfacN_t$. With negligible expected liquidation, the optimality index is simplified to include only the utilization error and the default term.



\begin{corollary}\label{corollary:optimality-index-optimizer}
Given a fixed set of parameters $\rolt, \robt, \sigma, \mu$ and assuming that $\userLiq{}{}(\liqthrshN_t, \colfacN_t) \approx 0$, the maximization problem of \optimalityNotation with respect to \(\colfacN_t\) can be formulated as follows:
\begin{align}
    \max_{\colfacN_t} \optimalityNotation &= \min_{\colfacN_t} \left(\frac{\rolt}{b + \frac{a}{\colfacN_t}} - \uoptimal\right)^2 + \gamma \frac{\rolt}{b + \frac{a}{\colfacN_t}} \Phi\left(\frac{\log(\colfacN_t) - \mu}{\sigma}\right) \\
    &\quad - \gamma\frac{\rolt}{b + \frac{a}{\colfacN_t}}\frac{\exp\left(\frac{\sigma^2}{2}+\mu\right)}{\colfacN_t}\cdot\Phi\left(\frac{-\mu+\log(\colfacN_t)-\sigma^2}{\sigma}\right)
\end{align}
where \( a \coloneqq \alpha \left(e^{\mu + \frac{\sigma^2}{2}} \frac{\Phi\left(\frac{-\mu - \sigma^2}{\sigma}\right)}{\Phi\left(\frac{-\mu}{\sigma}\right)} - 1\right) \) and \( b \coloneqq \alpha \robt + (1 - \alpha) \left(e^{\mu + \frac{\sigma^2}{2}} - 1\right) \).
\end{corollary}

\noindent We can derive this corollary by substituting the default, expected price fall, and price change terms from Lemma \ref{lemma:simplify-default-to-the-end} into the definition of \optimalityName stated in Definition \ref{def:optimality-index}.

\noindent\textbf{Estimator and optimizer construction} The estimator module, described in Algorithm \ref{alg:parameter-estimation-module}, uses a least squares estimator to analyze user behavior. It integrates outputs from the interest rate controller to determine the parameters \(\rolt\) and \(\robt\). Additionally, it learns the empirical price volatility \(\sigma\) and drift \(\mu\) from the recent price history.
 The optimizer component, detailed in Algorithm \ref{proc:optimizer}, utilizes these parameters to tackle the optimization problem presented in Corollary \ref{corollary:optimality-index-optimizer} and output the optimal collateral factor. 
To construct the estimator submodule, we model the lender's behavior using a linear regression approach, analogous to that of the borrower's behavior.
 This relationship is formulated as follows:
\begin{align}
    \frac{\lS{t+1} - \lt}{ \lt} &= \elasticityl \cdot \left( \rt\ut - \ut \,\poolDefault{}{}(\colfacN_t) - \rolt\right) + \varepsilon_t
\end{align}
where \(\poolDefault{}{}(\colfacN_t)\) is defined as per the simplifications in Lemma \ref{lemma:simplify-default-to-the-end}. The linear regression model for this behavior is represented by:
\begin{equation}\label{eq:linear-regression-lender}
\deltalmat = \plmat \cdot \thetalvec + \epsilonvec,
\end{equation}
with:
\begin{itemize}
    \item \(\deltalmat = \left[\frac{\lS{1} - \lS{0}}{ \lS{0}}, \frac{\lS{2} - \lS{1}}{ \lS{1}}, \ldots, \frac{\lS{t+1} - \lS{t}}{\lS{t}}\right]^T\) capturing the normalized changes in lenders' supply.
    \item \(\plmat = \begin{bmatrix}
        1 &  -\uS{0}\,\poolDefault{}(\colfacN_0) + \rS{0}\uS{0} \\
        1 &  -\uS{1}\,\poolDefault{}(\colfacN_1) + \rS{1}\uS{1} \\
        \vdots  & \vdots \\
        1  & -\ut\,\poolDefault{}(\colfacN_t) + \rt\ut
    \end{bmatrix}\) as the matrix of pool variables at each time step.
    \item \(\thetalvec = [-\elasticityl\,\rolt, \elasticityl]^T\) representing the parameter vector.
    \item \(\epsilonvec = [\varepsilon_{0}, \varepsilon_{1}, \ldots, \varepsilon_{t}]^T\) as the vector of noise terms.
\end{itemize}
Refer to Algorithm \ref{alg:parameter-estimation-module} and \ref{proc:optimizer} to find a detailed description of the planner.


\input{algorithms/optimizer}



\subsection{Adversarial Robustness}

In this section, we employ a gradient descent-based approach to perform robust linear regression, adapting the Torrent-GD method from the robust regression literature \cite{bhatia2015robust}. This method leverages the resilience of gradient descent to sparse corruptions and is highly efficient for large datasets, and helps us get an improved short-term algorithm for optimizing \AdversarialSus.

\noindent\textbf{Algorithm description}
The Torrent-GD modification to Algorithm 1 proceeds by updating the regression coefficients iteratively, reducing the influence of corrupt data points effectively. The update rule for the regression coefficients $\thetahatbvec$ at each iteration $t$ is given by:
\begin{equation}
    \thetahatbvec(t+1) = \thetahatbvec(t) - \kappa \nabla L_S(\thetahatbvec(t)),
\end{equation}

where $\kappa$ is the learning rate and $\nabla L_S(\thetahatbvec(t))$ represents the gradient of the loss function computed only over the subset of data points $S$ believed to be uncorrupted. This subset is dynamically determined in each iteration based on the residual errors.

\noindent\textbf{Gradient calculation}
The gradient of the loss function with respect to the regression coefficients is computed as:
\begin{equation}
    \nabla L_S(w) = (\pbmat{}^S)^T(\pbmat{})^S \thetahatbvec - \deltabmat_S,
\end{equation}
where $\pbmat{}^S$ and $\deltabmat_S$ are the features and responses of the uncorrupted subset, respectively.

\noindent\textbf{Active set update}
The active set $S$, which includes indices of the data points assumed to be uncorrupted, is updated using a hard thresholding operator that selects the points with the smallest residuals:
\begin{equation}
    S^{(t+1)} = \text{HT}\left(r^{(t)}, k\right),
\end{equation}
where $r^{(t)} = \deltabmat - \pbmat \thetahatbvec(t)$ represents the residuals at iteration $t$, and $k$ is a threshold parameter controlling the number of points included in $S$, and the hard thresholding operator $\text{HT}$ for a vector $v \in \mathbb{R}^n$ and a threshold $\tau$ is defined as follows:
\[
    \text{HT}(v_i,\tau) = 
    \begin{cases} 
    v_i & \text{if } |v_i| \geq \tau, \\
    0 & \text{otherwise}.
    \end{cases}
\]
This operation retains only the components of $v$ that are greater than or equal to the threshold $\tau$ in absolute value, effectively zeroing out smaller coefficients. 

\noindent\textbf{Convergence guarantees}
The convergence of Torrent-GD is guaranteed under conditions that the noise and corruptions are sparse \cite{bhatia2015robust}. The algorithm can tolerate arbitrary adversarial corruptions as long as $\beta < 50\%$.

\subsection{Blockchain implementation}
To implement this protocol on a blockchain, we can utilize an optimistic Rollup solution like Arbitrum or Optimism. The core idea behind these Rollups is that the computation is performed off-chain by a Rollup validator. Only the state update data is posted on the blockchain, allowing challengers to validate the state update with the list of user transactions and challenge the validators in case of discrepancies. Therefore, in an optimistic scenario, no computation is performed on-chain.

\noindent\textbf{Computational overhead} 
Our online algorithm and the parameter estimator modules each have a computational complexity of $\mathcal{O}(W)$, where $W$ is the sliding window used to collect and retain data for estimation. In practice, using $W \approx 50$ was sufficient in our implementation when the user's elasticity is not awfully low (An elasticity of approximately $10$ is sufficient when the noise standard deviation does not exceed $1$). 
The optimizer module, in general, can be computationally infeasible since its objective function is not necessarily convex. However, since the range of possible collateral factors is limited, we can discretize the possible values into approximately 100 levels and find the optimal collateral factor through brute force. The calculation of the objective function itself is not computationally intensive.\\
\noindent\textbf{Gas fee estimation}
Even the Roll-up solution might necessitate running the computation on-chain in case of a dispute, therefore we need to estimate the on-chain computation cost. To estimate the gas cost for updating the protocol's slow and fast parts, we consider the least squares estimator (LSE) update of Algorithm \ref{alg:interest-rate-controller} and \ref{alg:parameter-estimation-module}. This update requires approximately $12W$ multiplications and $10W$ additions, where $W$ is the length of the history window. Given the gas costs (5 gas per multiplication and 3 gas per addition based on \cite{eth-opcodes}), for $W=50$, we require $4500$ gas for multiplications and additions.
Additionally, storing new rows of $\pbmatt{}$ and $\deltabmat$ costs $20,000$ gas per 32-byte storage, totaling $40,000$ gas. Thus, the overall gas needed for the LSE update and storing new rows is $44,500$ gas.
The slow planner additionally needs to store a table of the optimality index values for different collateral factors and volatilities. Discretizing each into 100 and 10 values respectively, the storage cost for this table is $20,000 \times 1000 = 20,000,000$ gas. Additionally, storing a table of the CDF of Gaussian variables to calculate expected defaults and liquidations requires $20,000 \times 100 = 2,000,000$ gas for a reasonable discretization with $100$ points.
While more opcodes are involved in each interest rate and collateral factor update, they mainly consist of single arithmetic operations or multiple data storage, making the cost manageable.

\subsection{Limitations}
\noindent\textbf{User behaviour model}
We build upon a specific model of lender and borrower behavior, assuming that no single user controls a significant portion of the supply or demand. However, prior research has shown that this assumption might be unrealistic for many current platforms \cite{qin2021empirical, yaish2023suboptimality}. In these scenarios, rational user strategies would differ from those assumed in our model. While our model manages to capture the main factors driving both sides of the market in borrow-lending platforms, it fails to adapt to markets where a few entities control the majority of the funds.
Furthermore, we assume that liquidators are always available to liquidate positions when protocols permit. However, real data indicates that this is not always the case, especially when the collateral asset lacks liquidity, resulting in considerable slippage when reselling the collateral \cite{qin2021empirical}.

\noindent\textbf{Price distribution}
 Our choice of price model does not necessarily accurately describe the actual price distribution. However, as long as the price distribution belongs to a parameterized family of distributions with parameters that change slowly over time, the protocol can learn risk terms from user behavior, similar to our designed protocol. However, it may be challenging to analytically infer the price distribution parameters from user actions or to analytically relate the price distribution to key risk metrics, such as expected default or price fall. 

 \noindent\textbf{Risk neutralily}
 In defining our utility functions, we assume users are risk-neutral and perceive their utility as the sum of profit minus expected risk, accurately calculating expectations over the price distribution. However, this may not be realistic. Our protocol learns the projection of user behavior onto our specific linear utility function and identifies parameters that best describe this behavior. For more complex user behavior, neural networks can be used to learn a vector representation of user behavior end-to-end. These user representation vectors are then fed to the interest rate controller and collateral factor planner, which are replaced by Deep Reinforcement Learning agents. These agents learn the optimal strategy based on feedback from their reward function. A key challenge is designing reward functions that best meet the protocol's needs.

 \noindent\textbf{Adversarial robustness}
 The adversarial resistance model we use for the LSE algorithm may be inadequate for permissionless blockchains. In such blockchains, any participant, including miners who organize and submit transactions, can act as adversaries. These adversaries can run the regression algorithm off-chain with different sets of transactions and find the set and the order that benefits them the most. Our current notion of adversarial robustness only protects against adversaries who control less than 50\% of the funds and occasionally send transactions that deviate from the system's assumed supply and demand dynamics i.e., producing outlier data points.

\noindent\textbf{Single equilibrium point}
Our borrower behavior model assumes that there is always a single, unique interest rate that all borrowers are willing to pay, and that this equilibrium interest rate is known to all borrowers with some noise. However, in reality, this assumption may not hold true. Different groups of borrowers may have varying perceptions of the interest rate they are willing to pay. As a result, the system could have multiple equilibrium points, each attracting a different subgroup of borrowers.


%% file: algorithms/bestimator.tex
\begin{algorithm}
\begin{algorithmic}[1]\caption{Interest rate controller, utilizing a least squares optimization approach}
\label{alg:interest-rate-controller}
    \State Initialize: \(t \gets 0\), \(\delta \gets\) stability threshold, \(t_{\text{sleep}} \gets\) sleep time, \(\nu\) exploration probability
    \While{True} 
        \If{ $\frac{\bt - \bS{t-1}}{\bS{t-1}} < \delta$}
        \State Sleep for $t_{\text{sleep}}$
        \State Reset matrices $\deltabmat$ and $\pbmat$
        \Else{}
        
        \State Add the new row $[1,\rS{t-1}]$ to \(\pbmatt{t-2}\) to construct \(\pbmatt{t-1}\) and \State Add the new column $[\frac{\bt - \bS{t-1}}{ \bS{t-1}}]$ to \(\deltabmat\)
        \State Perform least squares estimation to find 
        \(\thetahatbvec \gets ({(\pbmatt{t-1})}^T\pbmatt{t-1})^{-1} \,(\pbmatt{t-1})^T\, \deltabmat\)
        \State Parse \thetahatbvec as $[\hat{\elasticityb}\hat{r}^*, -\hat{\elasticityb}]$ and extract $\hat{r}^*$ and set $\rt = \hat{r}^*$
        \State With probability $\nu$, choose a random $\rt \in [r_\text{min}, r_\text{max}]$
        \EndIf
        \State $t \gets t+1$
    \EndWhile
    \State \textbf{end algorithm}
\end{algorithmic}
\end{algorithm}

%% file: algorithms/optimizer.tex
\begin{algorithm}[!t]
\caption{$\rolhat, \robhat$ - Estimator, auxiliary to the optimizer procedure}\label{alg:parameter-estimation-module}
\begin{algorithmic}[1]
    \State Initialize: \(t \gets 1\), $\thetahatlvec^0 \gets \mathbf{0}$, and read $\lS{0}, \uS{0}, \rS{0}, \colfacN$ from the pool, \(\delta_{l} \gets\) stability threshold
    \State Initialize: \(t_{\text{sleep}} \gets\) sleep time, $\delta_{\theta} \gets$ least square convergence threshold, \(T_{\text{optimizer}} \gets\) the minimum time interval between successive executions of the optimizer. \(\alpha \gets \) fraction of first type borrowers
    \While{True}
        \If{ $\frac{\lt - \lS{t-1}}{\lS{t-1}} < \delta_l$}
            \State Reset $\deltalmat, \plmat $
            \State Sleep for $t_{\text{sleep}}$
        \Else
            \State Calculate $\sigma, \mu$ empirically from the recent price history and use them to calculate $\poolDefault{}{}(\colfacN_t)$
            \State Add the new row $[1,-\uS{t-1}\,\poolDefault{}(\colfacN_t)+ \rS{t-1}\uS{t-1}]$ to \(\plmat\) and $[\frac{\lt - \lS{t-1}}{ \lS{t-1}}]$ to \(\deltalmat\)
            \State Perform least squares estimation to find 
            \(\thetahatlvec^t \gets ({\plmat}^T\plmat)^{-1} \,\plmat^T\, \deltalmat\)
            \State Parse $\thetahatlvec^t$ as $[-\hat{\elasticityl}\,\rolhat\,,\,\hat{\elasticityl}]^T$ and extract $\rolhat$
            \State Read the latest $\hat{r}^*$ from Algorithm \ref{alg:interest-rate-controller}, and extract $\robhat$ as follows: 
            \State\(\alpha\,\robhat \gets \hat{r}^*  - {\poolDefault{}}(\colfacN_t) - \frac{\alpha}{\colfacN_t} \left(e^{\mu + \frac{\sigma^2}{2}} \frac{\Phi\left(\frac{-\mu - \sigma^2}{\sigma}\right)}{\Phi\left(\frac{-\mu}{\sigma}\right)} - 1\right) - (1 - \alpha) \left(e^{\mu + \frac{\sigma^2}{2}} - 1\right)\)
            
            \If{$|\thetahatlvec^t-\thetahatlvec^{t-1}| < \delta_{\theta}$}
                \State $\colfacN \gets$ \Call{Optimizer}{$\rolhat, \robhat, \sigma$}
                \State Reset $\deltalmat, \plmat $
                \State sleep for \(T_{\text{optimizer}}\)
            \EndIf
        \EndIf
        \State $t \gets t + 1$
    \EndWhile
    \State \textbf{end algorithm}
\end{algorithmic}
\end{algorithm}

\begin{algorithm}[th]
\caption{Optimizer module}\label{proc:optimizer}
\begin{algorithmic}[1]
    \Procedure{Optimizer}{$\rolhat, \robhat, \sigma$}
        \State Initialize: \(\alpha \gets \) fraction of first type borrowers, $\uoptimal \gets$ desired utilization level, initial guess for collateral factor $c(0) \overset{\text{U}}{\sim} [0,1]$, $ \gamma \gets$ default regularization factor, $\kappa \gets$ learning rate, $i \gets 0$ the gradient descent iterator, $\delta \gets$ gradient descent stop theshold.
        \State Set $a = \alpha \left(\exp\left(\mu - \frac{\sigma^2}{2}\right)-1\right)$ and $b = \alpha \,\robhat  + (1 - \alpha) \left(\exp\left(\mu + \frac{\sigma^2}{2}\right) - 1\right)$
        \State $\Psi(c)\coloneqq\left(\left(\frac{\rolhat}{b + \frac{a}{\colfacN}}  - \uoptimal\right)^2 + \gamma \, \frac{\rolhat}{b + \frac{a}{\colfacN}} \, \Phi\left(\frac{\log(\colfacN) - \mu}{\sigma}\right) -  \lambda\frac{\rolt}{b + \frac{a}{\colfacN}}\frac{\exp{(\frac{\sigma^2}{2}+\mu)}}{\colfacN}\cdot\Phi\left(\frac{-\mu+\log(\colfacN)-\sigma^2}{\sigma}\right)\right)$
        
        \While{$|\Psi\left(\colfacN(i)\right)- \Psi\left(\colfacN(i-1)\right)|> \delta$}
        \State $i \gets i + 1$
            
            \State $\colfacN(i) \gets \colfacN({i-1}) - \kappa \, \frac{d\, \Psi(c)}{d\,\colfacN}|_{\colfacN=\colfacN(i-1)}$
        \EndWhile
    
        \State \textbf{return} \( c(i) \)
    \EndProcedure
\end{algorithmic}
\end{algorithm}

%% file: evaluation.tex
\section{Evaluation}\label{sec:eval}
In this section, we test the interest rate controller and collateral factor planner described in section \ref{sec:protocol-design-guarantees} and demonstrate their robustness to the change of market and user behaviours, moreover, we compare our interest rate controller with that of the baseline.

\subsection{Interest rate controller}
\begin{figure}[thbp]
    \centering
    \begin{subfigure}[t]{0.485\textwidth}
        \centering
        \includegraphics[width=\textwidth]{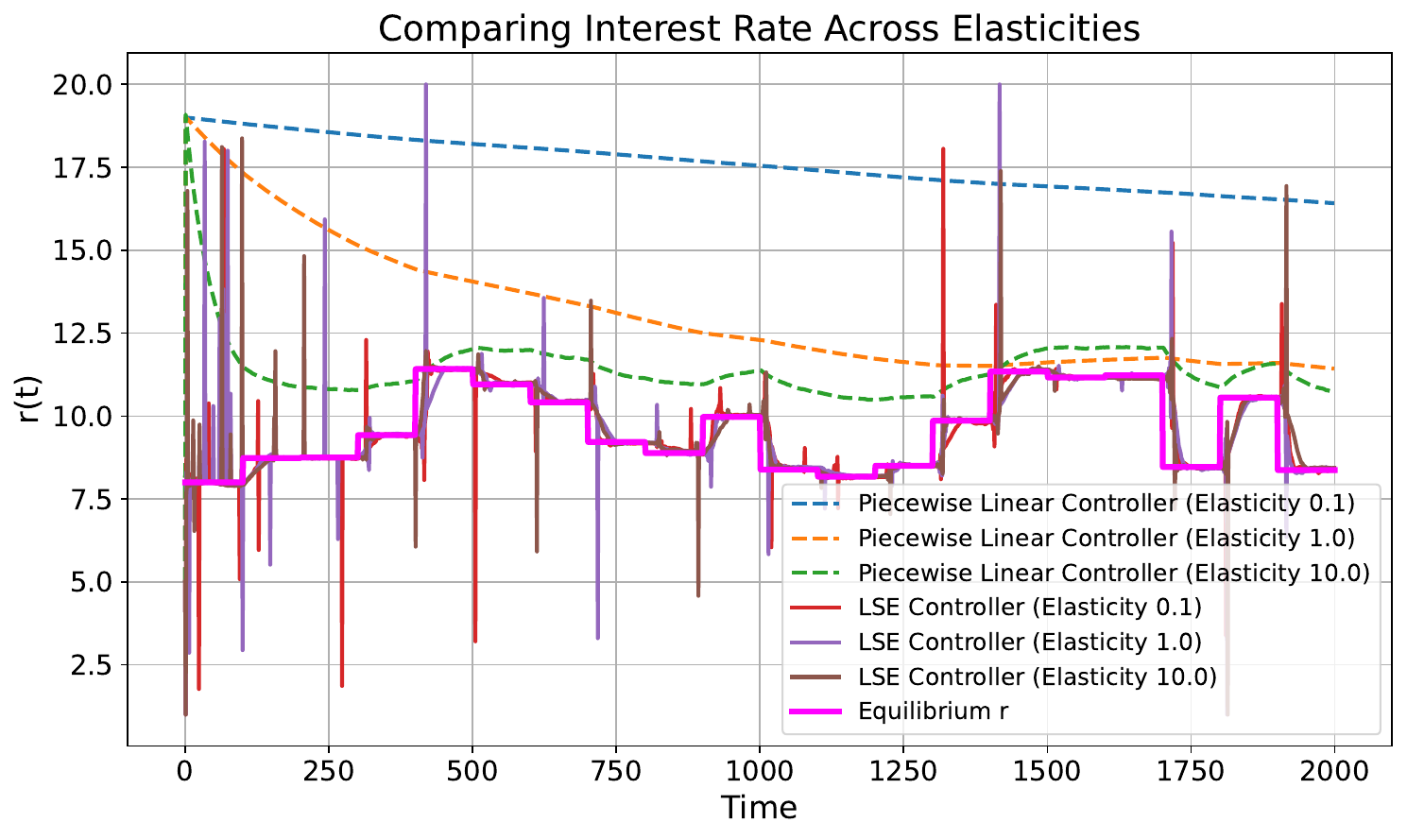}
        \caption{The LSE-based controller adapts to the new equilibrium interest rate quickly. In contrast, the piecewise linear interest rate curve fails to track the equilibrium interest rate when the borrower's elasticity is low.}
        \label{fig:r_comparison}
    \end{subfigure}
    \hfill
    \begin{subfigure}[t]{0.485\textwidth}
        \centering
        \includegraphics[width=\textwidth]{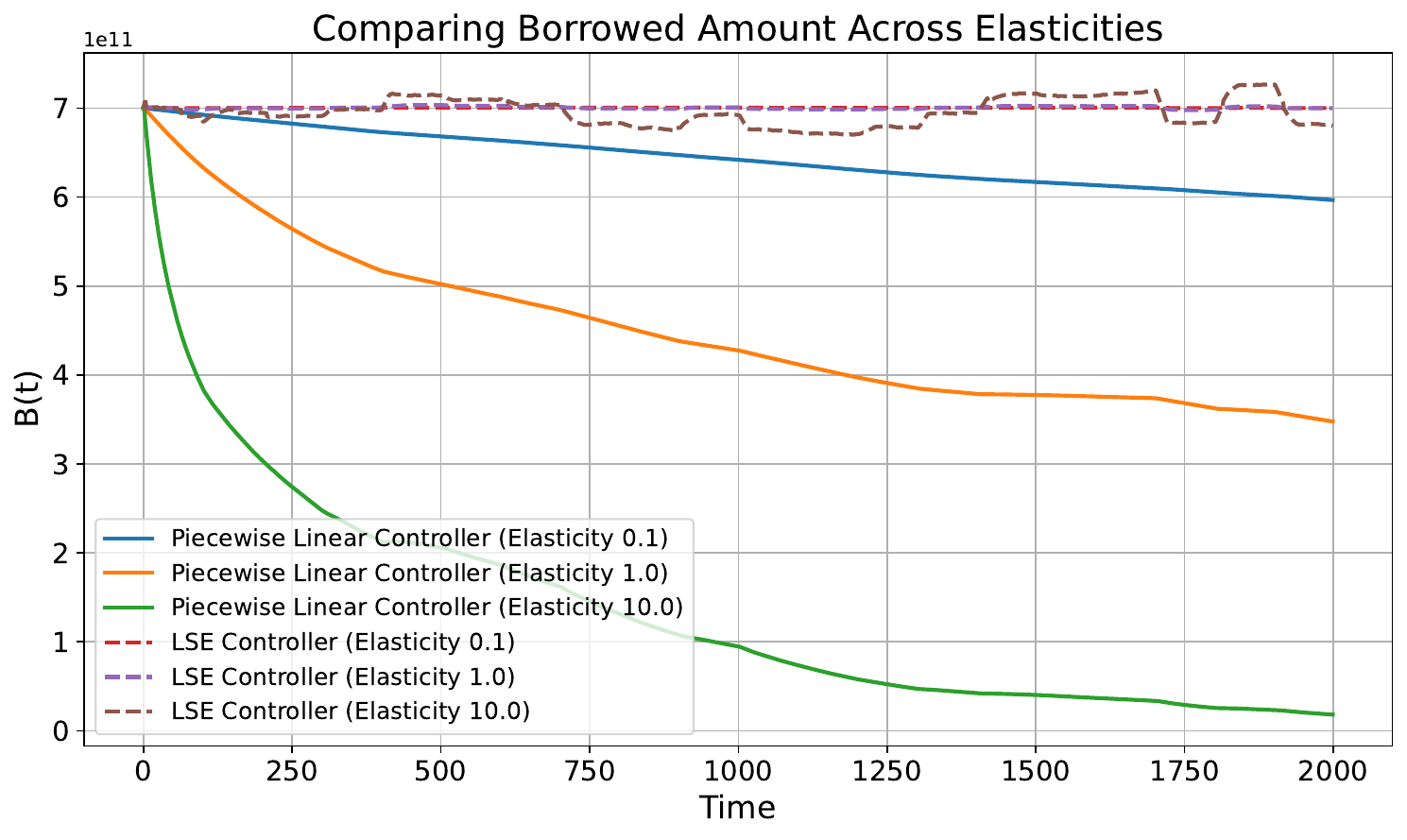}
        \caption{The LSE controller prevents the exit of borrower capital by setting a competitive interest rate, unlike the baseline controller.}
        \label{fig:b_comparison}
    \end{subfigure}
    \caption{Comparing the LSE-based interest rate controller and the piecewise linear curve.}
    \label{fig:comparison}
\end{figure}

\noindent\textbf{Set-up} In this experiment, we start a borrow-lending pool with an initial borrow supply of $7 \times 10^{11}$ and an initial lend supply of $10^{12}$. Both the elasticity of borrowing and lending are set to 50, and the standard deviation of the borrowed and supply dynamic noise is 0.1. We assume very low price volatility, as it does not affect the determination of the equilibrium interest rate in this experiment. Every 100 time slots, we change $\robt$ and allow the interest rate controller to adjust to the new equilibrium interest rate based on borrowers' behavior. During the exploration phases, the interest rate is randomly selected between $r_{min}=1$ and $r_{max}=20$ (outliers in figure \ref{fig:r_comparison}). 

\noindent\textbf{Tracking The equilibrium interest rate} As shown in Figure \ref{fig:r_comparison}, the LSE-based controller adapts to the equilibrium interest rate consistently across different user elasticities with a nearly identical convergence rate, aligning with our theoretical guarantees. In contrast, the baseline controller, which sets the interest rate as a piecewise linear function of utilization, heavily relies on elasticity. It struggles to adapt to the new equilibrium interest rate in low elasticity scenarios and performs slightly better as elasticity increases.

\noindent\textbf{Borrowers drain due to wrong rate} While the performance of the baseline controller improves as borrower's elasticity increases, the consequences of misadjustment are more severe, causing significant borrower capital to leave the system when the interest rate is too high and to flood the system when it is too low.
Figure \ref{fig:b_comparison} illustrates how the borrowed value changes with market conditions. The LSE-based controller consistently finds the stable interest rate, resulting in minimal market disruption whenever changes occur. In contrast, the baseline controller's inability to quickly find the correct rate causes substantial borrower exit from the system. This issue worsens with increased borrower elasticity due to higher repayment rates. Thus, even in high elasticity markets, the baseline controller is ineffective in preventing excessive capital inflows or outflows due to interest rate misadjustments.

\subsection{Collateral factor planner}

\begin{figure}[htbp]
    \centering
    \includegraphics[scale=0.5]{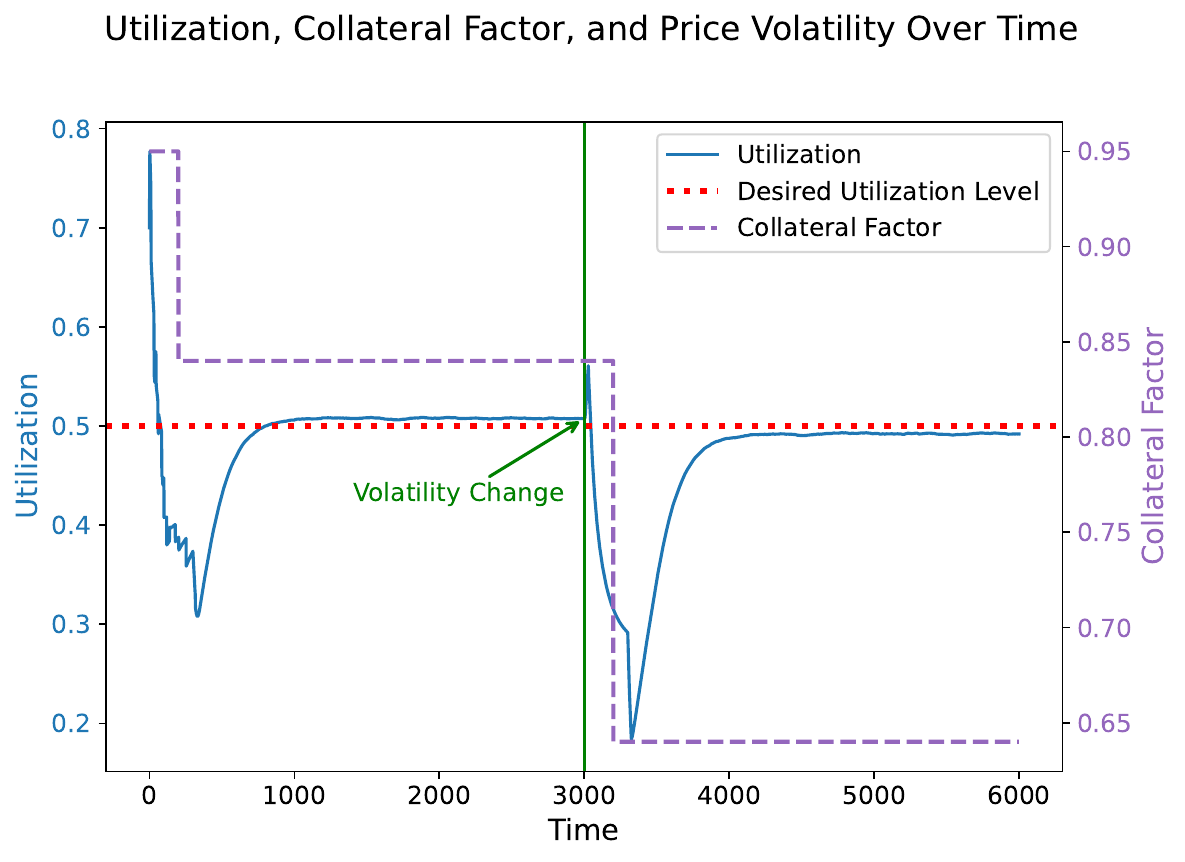} 
    \caption{}
    \label{fig:u}
\end{figure}

We conducted an experiment to evaluate the performance of our collateral factor planner. The \optimalityName is defined as in Definition \ref{def:optimality-index} with \(\lambda = 0\) and \(\uoptimal = 0.5\). Initially, the system starts with an unoptimized collateral factor of \(\colfacN = 0.95\). At timeslot \(t = 200\), the optimizer activates and sets a new collateral factor of \(\colfacN = 0.84\), which adjusts the utilization to the desired level of \(0.5\) soon after that (the pace of reaching the equilibrium is a function of lenders' elasticity).
At timeslot \(t = 3000\), a change in price volatility disrupts the system. By timeslot \(t = 3200\), the estimator accurately detects the new volatility and triggers the optimizer. The optimizer then sets a new collateral factor of \(\colfacN = 0.64\), which stabilizes the utilization around \(0.5\) once again. This entire process is automated. The resulting utilization curve is shown in Figure \ref{fig:u}.

%% file: system-design.tex

%% file: conclusion.tex
\section{Conclusion and discussion}

\noindent\textbf{Modelling deFi borrow-lending user behavior} 
In this paper, we present a first-principles model of the behavior and incentives of borrowers and lenders in a DeFi market. We consider their alternative strategies and analyze how to achieve market equilibrium in the presence of price volatility. We mention empirical evidence on the validity of our model.

\noindent\textbf{Two-level protocol design}
We propose a data-driven, borrow-lending protocol that sets the interest rate and over-collateralization ratio adaptively. By monitoring user reactions and learning from their behavior, our protocol determines a competitive interest rates and optimal collateral factors. The protocol consists of two components, 1) Fast interest rate controller: This component reacts online to user behavior, ensuring competitive interest rates and preventing over- or underpaying users. It has theoretical guarantees for fast convergence to the equilibrium interest rate. 2) Slow collateral factor planner: This component uses accurate market condition estimates to adjust the collateral factor, maintaining utilization at a desired level while controlling default risk. Overall, our protocol ensures rapid convergence to the equilibrium interest rate and optimal tuning of the collateral factor to achieve a desired equilibrium.

\noindent\textbf{Implementation}
We implement our protocol and test its performance using simulated users, including uninformed ones. We compare our protocol with a baseline that uses piecewise linear functions to set interest rates based on utilization. Our protocol demonstrates superior performance compared to the baseline in practice.

%% file: appendix.tex
\begin{appendices}
\newpage

\section{Supplementary explanation on the protocol logic}\label{app:protocol-logic}
\subparagraph{A Comment on Handling Defaults}
Unlike most lending platforms, our protocol does not use a reserve pool to cover defaults. Traditional DeFi lending protocols charge borrowers higher interest rates to build a reserve for defaults. Instead, our protocol directly deducts defaults from lenders' revenue each timeslot, removing risk for the protocol. Both approaches manage default risk: traditional models collect reserves continuously, while our model applies the adjustment explicitly. Our approach can still be used if a protocol aims to mitigate risk by collecting a reserve pool. The expected default we calculate can serve as a metric to charge borrowers more, thus avoiding direct deductions from lenders' revenue.

\section{Supporting evidence for our user model}
In this section, we discuss our lender and borrower model in more detail and explain why we believe it closely reflects Defi borrow-lending platform user behavior. Moreover, we provide evidence from previous works and technical reports from Aave and Compound to support the variables introduced in our model. Additionally, we provide empirical estimates of these variables based on real-world data \cite{saengchote2023decentralized, aave-report-2023}.

\subsection{Lender}\label{appendix:subsec:competition}
\subparagraph{What does \(\rolt\) correspond to in the real world?} In our model, we represent competition with the external market using two parameters: \(\rolt\) for lenders and \(\robt\) for borrowers. For lenders, although there is no explicit risk-free interest rate outside the market representing \(\rolt\), they have various alternative strategies if they decide to allocate part of their portfolio to the asset \assetOne. These alternatives include:

\begin{itemize}
    \item Participating in yield farming by providing liquidity to decentralized exchanges like Uniswap or SushiSwap.
    \item Staking tokens in Proof-of-Stake networks like Ethereum 2.0.
    \item Using automated investment platforms like Yearn Finance to optimize yields across different protocols.
    \item Depositing stablecoins like USDC, DAI, or USDT in high-yield savings accounts on platforms like Anchor Protocol.
    \item Participating in stablecoin-specific liquidity pools on Curve Finance.
    \item Using fixed-rate lending protocols like Notional Finance for predictable returns.
\end{itemize}

Different lenders may have varying levels of knowledge about these alternatives and may or may not consider the benefits of native tokens from platforms like Compound and Aave. Regardless, each lender expects a minimum interest rate \(\rolt\) below which they will withdraw their funds. This \(\rolt\) can be zero or even negative for some users due to the other types of rewards that these Defi platforms provide such as interest rate on the native token. The \(\rolt\) in our model can be interpreted as the weighted average of \(\rolt\) values for all users, weighted by their deposit amounts.

We assume lenders are incentivized by a linear function of the interest rate. However, this might not always be accurate, as lenders could expect higher future interest rates and tolerate current lower rates to avoid transaction fees. Since we allow \(\rolt\) to change arbitrarily over time, our model can accommodate these affects. For instance, high transaction fees disincentivizing withdrawals will result in a lower \(\rolt\) for those lenders, which will be reflected in the lender's behavior.

\subparagraph{Elasticity} Aave and Compound have conducted empirical studies on the elasticity of lenders and borrowers by analyzing their reactions to changes in interest rate curves and other parameters. These studies typically measure elasticity by observing the quantity of assets moved by users following any change and the speed of their reactions. This data is then used to perform counterfactual analyses of proposed changes \cite{aave-report-2023}.

In our model, we represent users' elasticity with parameters $\elasticityb$ and $\elasticityl$, assuming these parameters can change over time but not rapidly (only on the order of $\Tmarket$). This approach aligns with previous findings that show users' elasticity can vary over time and in different occasions \cite{aave-report-2023}, with some users even displaying zero elasticity in certain cases. By making the rate of borrow-lend change a linear function of elasticity, our quantitative model effectively incorporates these dynamics.

\subsection{Borrower}\label{appendix:subsec:Tuser}\label{appendix:subsec:borrower-behaviour}


\subparagraph{ Recursive Borrowing and Leveraged Yield Farming} Saengchote notes that a significant portion of loan demand on the Compound protocol comes from a practice called \emph{leveraged yield farming }\cite{saengchote2023decentralized}. In this strategy, users repeatedly borrow and deposit the same asset to earn native token rewards, which are shared among all active lenders and borrowers by the protocol. These rewards can sometimes be so high that they offset the interest paid by borrowers, resulting in a positive net value for them. However, our model does not consider this strategy. Instead, we focus on user incentives in a simple borrow-lending market, excluding protocol-specific factors like borrowing or deposit rewards.
Now we provide more details and explanations on our borrower model:

\subsubsection{Financing Borrowers}
 We represent the value these borrowers receive from the borrowed asset as \(\robt\). Similar to \(\rolt\), \(\robt\) is measured as an interest rate and helps assess the profit that \protocol provides by offering liquidity.

These borrowers are exposed to the risk of locked collateral, meaning that if the collateral were not locked, they could sell it at the first sign of price decay. With the collateral locked in the pool, they cannot do this, so any price drop must be accounted for in their utility function as the opportunity cost of having their collateral locked.

We define the utility function for a typical borrower of this type, indexed by \( i \), who holds \(\ct(i)\) collateral:

\begin{equation}\label{eq:borrower-one-effective-r}
\begin{aligned}
    \utilitybOne(i) \coloneqq &\robt - \rt + \E{\userDefault{t}{i}(\pS{t+1})} \\
    &- \E{\userLiq{t}{i}(\pS{t+1})} \cdot \liqIncntive_t \\
    &+ \ct(i) \cdot \E{\indicator{\pS{t+1} - \pt < 0}(\pS{t+1} - \pt)}
\end{aligned}
\end{equation}

This utility function can be broken down as follows:
1. Inherent Value (\(\robt\)): Represents the annualized value of one unit of liquidity to the borrower.
2. Interest Rate (\(-\rt\)): The annualized interest rate the borrower must pay.
3. Default Value (\(\E{\userDefault{}{i}(\pS{t+\T})}\)): If default occurs, the borrower can escape with the defaulted value.
4. Liquidation Cost (\(-\E{\userLiq{}{i}(\pS{t+1})} \cdot \liqIncntive_t\)): Represents the liquidation fee that is deducted from the user's collateral if liquidation occurs.
5. Opportunity Cost (\(\ct(i) \cdot \E{\indicator{\pS{t+1} - \pt < 0}(\pS{t+1} - \pt)}\)): Accounts for the opportunity cost of having collateral locked in the contract when the price drops.

This opportunity cost is based on an alternative strategy where the collateral could be sold at the first sign of price decline. Note that we only subtract this penalty term if the price drops and don't add a corresponding gain for a price rise since this type of borrower already holds a profile of \(\assetTwo\). Thus, their profit from price appreciation remains unaffected by the protocol but differs in the event of a price fall due to limited reaction speed.

\noindent\textbf{Empirical Validation} The empirical results found by Saengchote validate our model's inclusion of a punishment term for expected price falls in the borrower's utility function. They analyzed borrow positions from the Compound protocol from July 2019 to June 2020 and observed that loan demand has a negative correlation with the past 30-day standard deviation of ETH for USDC, USDT, and TUSD. This correlation is stronger in stablecoin-to-cryptocurrency pools compared to cryptocurrency-to-cryptocurrency pools, which experience less relative volatility \cite{saengchote2023decentralized}.

\subsubsection{Leveraged Trading Borrowers}

Each borrower of this type seeks to open a leveraged position in \(\assetTwo\) worth $\$1$ at time \( t \). Their strategy is as follows: they borrow \(\frac{1}{\pt}\) units of \(\assetTwo\) (worth $\$1$) from an external provider \(\mathcal{Z}\), then add additional collateral in the amount of \(\$ \left(\frac{1}{\colfacN_t} - 1\right)\) to secure a loan of 1 unit of \(\assetOne\) from \protocol. Since the maximum loan-to-value ratio allowed by \protocol is \(\colfacN_t\), the borrower must provide collateral worth \(\$ \frac{1}{\colfacN_t}\), equivalent to \(\frac{1}{\colfacN_t \cdot \pt}\) units of \(\assetTwo\).

Next, they immediately exchange the borrowed $1$ \(\assetOne\) for \(\frac{1}{\pt}\) units of \(\assetTwo\) through an exchange to repay their debt to \(\mathcal{Z}\). This results in a leveraged position of \(\frac{1}{\pt}\) units of \(\assetTwo\) in \protocol, from which they will profit if the price of \(\assetTwo\) rises. Their rationale is that \(\assetTwo\) will appreciate in value, making their position in \protocol advantageous. These borrowers have been studied extensively in \cite{szpruch2024leveraged}.

The utility function for this group is defined as follows:

\begin{equation}\label{eq:borrower-two-effective-r}
    \utilitybTwo(i) = -\rt + \E{\userDefault{t}{i}(\pS{t+1})} - \E{\userLiq{t}{i}(\pS{t+1})} \cdot \liqIncntive_t + \E{\frac{\pS{t+1} - \pt}{\pt}}
\end{equation}
Where all the expectations are over $\pS{t+1}$.

The utility function of these borrowers resembles that of the first group but lacks the \(\robt\) term because these borrowers do not derive inherent value from \(\assetOne\). Instead, they immediately swap it for \(\assetTwo\), as their primary interest in \protocol is to take a long position on \(\assetTwo\).

Unlike the first group, these borrowers receive positive utility from an increase in \(\assetTwo\)'s price because they did not own this amount of \(\assetTwo\) before engaging with \protocol. They can only invest in \(\frac{1}{\pt}\) units of \(\assetTwo\) after joining \protocol. While the first type of borrowers already holds a profile of \(\assetTwo\) and therefore doesn't gain utility from price increases, the second group can only establish this profile through \protocol, so all gains from investing in \(\assetTwo\) should be reflected in their utility function to capture the full profit and risk.

\section{Proofs}
\subsection{Proof of lemma \ref{lemma:collateral-strategy}}

\begin{proof}

We denote user \( i \)'s loan-to-value ratio at time \( t \) as \( ltv_t^i \), and assume \( ltv_t^i < \colfacN_t \). Since the borrowed amount per user is always 1 unit, we have:

\[
\frac{1}{\ct(i) \pt} = ltv_t^i \quad \Rightarrow \quad \ct(i) = \frac{1}{\pt \cdot ltv_t^i}.
\]

We can rewrite the utility function for a type 2 borrower by substituting \(\ct(i)\), yielding:

\[
\userDefault{}{i}(\pS{t+1}) = \max\{0, \bt(i) - \ct(i) \pS{t+1}\} = \max\{0, 1 - \frac{\pS{t+1}}{\pt \cdot ltv_t^i}\}.
\]

This allows us to express the type 2 borrower's utility function as:

\[
\utilitybTwo(i) = -\rt + \max\{0, 1 - \frac{\pS{t+1}}{\pt \cdot ltv_t^i}\} - \E{\frac{\pS{t+1} - \pt}{\pt}}.
\]

Since only the default term depends on \( ltv_t^i \), a rational user will:

\[
ltv_t^i = \arg \max_{ltv} \E{\max\{0, 1 - \frac{\pS{t+1}}{\pt \cdot ltv}\}}.
\]

Breaking down the expression:

\[
ltv_t^i = \arg \max_{ltv} \prob{\pS{t+1} < \pt \cdot ltv} \cdot \E{1 - \frac{\pS{t+1}}{\pt \cdot ltv} \mid \pS{t+1} < \pt \cdot ltv}.
\]

Expanding further:

\[
ltv_t^i = \arg \max_{ltv} \int_{0}^{\pt \cdot ltv} \left(1 - \frac{p}{\pt \cdot ltv}\right) f(p)\, dp.
\]

This simplifies to:

\[
ltv_t^i = \arg \max_{ltv} \prob{\pS{t+1} < \pt \cdot ltv} - \frac{1}{\pt \cdot ltv} \int_{0}^{\pt \cdot ltv} p f(p)\, dp = \colfacN_t.
\]

The expression above is the sum of two functions that both increase with \( ltv \), for any known \(\pt\) and price distribution of \(\pS{t+1}\). Thus, the optimal \( ltv \) is the maximum allowed by the protocol, \(\colfacN_t\).

Next, we analyze the utility function for a type 1 borrower. We write the utility function as a function of \( ltv_t^i \), ignoring the liquidation cost term:

\[
\utilitybOne(i) = \robt - \rt + \E{\max\{0, 1 - \frac{\pS{t+1}}{\pt \cdot ltv_t^i}\}} + \frac{1}{\pt \cdot ltv_t^i} \cdot \E{\indicator{\pS{t+1} - \pt < 0} (\pS{t+1} - \pt)}.
\]

Since the first two terms are independent of \( ltv_t^i \), a rational type 1 borrower will:

\[
ltv_t^i = \arg \max_{ltv} \E{\max\{0, 1 - \frac{\pS{t+1}}{\pt \cdot ltv}\}} + \frac{1}{\pt \cdot ltv} \cdot \E{\indicator{\pS{t+1} - \pt < 0} (\pS{t+1} - \pt)}.
\]

The first term is identical to that analyzed for type 2 borrowers. The second term is also an increasing function of \( ltv \) because \(\E{\indicator{\pS{t+1} - \pt < 0} (\pS{t+1} - \pt)} < 0\). Therefore, the overall function is increasing in \( ltv \), and the optimal \( ltv \) is the maximum allowed by the protocol, \(\colfacN_t\).

\end{proof}

\subsection{proof of lemma \ref{lemma:simplify-default-to-the-end} }

\paragraph*{Expected Default}

\begin{proof}
We first simplify:

\begin{align}\label{eq:pool-default-simplify}
     \poolDefault{t}(\pS{t+1})&= \frac{1}{\lt}\sum_{i \in \mathrm{Borrowers}} \max\left\{0, \bt(i) - \ct(i) \cdot \pS{t+1}\right\} \\ 
    &\overset{\text{Lemma \ref{lemma:collateral-strategy}}}{=} \frac{1}{\lt}\sum_{i \in \mathrm{Borrowers}} \max\{0, 1 - \frac{\pS{t+1}}{\colfacN_t \cdot \pt}\} \\
    &= \label{eq:simplified-default}\frac{\bt}{\lt} \cdot \max\{0, 1 - \frac{\pS{t+1}}{\colfacN_t \cdot \pt}\}=\ut \,\max\{0, 1 - \frac{\pS{t+1}}{\colfacN_t \cdot \pt}\}
\end{align}
And similarly:
\[\userDefault{t}{i} = \max\{0, 1 - \frac{\pS{t+1}}{\colfacN_t \cdot \pt}\}\]

Let \( X = \frac{\pS{t+1}}{\pt} \) be a random variable where \( \log(X) \) follows a normal distribution with mean \( \mu \) and variance \( \sigma^2 \), making \( X \) a log-normally distributed variable. The probability density function (PDF) of \( X \) is given by:
\[
f_X(x) = \frac{1}{x\sigma\sqrt{2\pi}} \exp\left(-\frac{(\log(x) - \mu)^2}{2\sigma^2}\right), \quad x > 0.
\]
We seek to compute the expectation: (for ease of notation we use $\colfacN$ instead of $\colfacN_t$ )
\[
\mathbb{E}\left[\max\left\{0, 1 - \frac{X}{\colfacN}\right\}\right].
\]
This expectation can be split into two integrals based on the value of \( X \):
\[
\mathbb{E}\left[\max\left\{0, 1 - \frac{X}{\colfacN}\right\}\right] = \int_0^{\colfacN} \left(1 - \frac{x}{\colfacN}\right) f_X(x) \, dx.
\]

Let us define \( u = \frac{x}{\colfacN} \), then \( x = u\colfacN \) and \( dx = \colfacN du \). Substituting and adjusting the limits gives:
\[
\int_0^1 \left(1 - u\right) f_X(u \colfacN) \colfacN \, du.
\]

Using the transformation \( x = u \colfacN \) in the PDF of \( X \), we can write:
\[
f_X(u \colfacN) = \frac{1}{u \colfacN \sigma \sqrt{2\pi}} \exp\left(-\frac{(\log(u \colfacN) - \mu)^2}{2\sigma^2}\right).
\]

Thus, the expectation becomes:
\[
\mathbb{E}\left[\max\left\{0, 1 - \frac{X}{\colfacN}\right\}\right] = \int_0^1 (1 - u) \frac{1}{u \sigma \sqrt{2\pi}} \exp\left(-\frac{(\log(u) + \log(\colfacN) - \mu)^2}{2\sigma^2}\right) du.
\]

we explore the steps involved in simplifying this integral, which incorporates the probability density function of a log-normally distributed variable, transformed via scaling.

Given the complex integral:
\[
\int_0^1 (1 - u) \frac{1}{u \sigma \sqrt{2\pi}} \exp\left(-\frac{(\log(u) + \log(\colfacN) - \mu)^2}{2\sigma^2}\right) du,
\]
we aim to simplify it using substitution and properties of the normal distribution.

Transform the integral by substituting \( v = \log(u) \), hence \( u = e^v \) and \( du = e^v dv \). This converts the integral into:
\[
\int_{-\infty}^0 (1 - e^v) \frac{1}{\sigma \sqrt{2\pi}} \exp\left(-\frac{(v + \log(\colfacN) - \mu)^2}{2\sigma^2}\right) dv = \Phi\left(\frac{\log(\colfacN) - \mu}{\sigma}\right)
\]

Considering \( v = \log(u) \), and \( v + \log(\colfacN) - \mu \) follows a normal distribution, we leverage the cumulative distribution function (CDF) of the normal distribution. This CDF evaluation simplifies the integral:
\[
\Phi\left(\frac{\log(\colfacN) - \mu}{\sigma}\right) - \frac{\exp{(\frac{\sigma^2}{2}+\mu)}}{\colfacN}\cdot\Phi\left(\frac{-\mu+\log(\colfacN)-\sigma^2}{\sigma}\right)
\]


\end{proof}
\paragraph*{Expected Price fall}
\begin{proof}

Let \(X = \log\left(\frac{p_{t+1}}{p_t}\right)\). Then \(X \sim \mathcal{N}(\mu, \sigma^2)\).
Since \(X = \log\left(\frac{p_{t+1}}{p_t}\right)\), we have \(p_{t+1} = p_t e^X\).
The condition \(p_{t+1} < p_t\) is equivalent to \(e^X < 1\), which implies \(X < 0\).
\[
   \mathbb{E}[p_{t+1} - p_t \mid p_{t+1} < p_t] = \mathbb{E}[p_t e^X - p_t \mid X < 0] = p_t \mathbb{E}[e^X - 1 \mid X < 0]
\]
   \[
   \mathbb{E}[e^X - 1 \mid X < 0] = \mathbb{E}[e^X \mid X < 0] - 1
   \]
   For a normal random variable \(X \sim \mathcal{N}(\mu, \sigma^2)\), the conditional expectation \(\mathbb{E}[e^X \mid X < 0]\) can be found using the fact that \(e^X\) is log-normally distributed.

The expectation of a truncated log-normal variable \(e^X\) given \(X < 0\) can be derived using the cumulative distribution function (CDF) and the probability density function (PDF) of the normal distribution. Specifically,
\[
\mathbb{E}[e^X \mid X < 0] = \frac{\int_{-\infty}^{0} e^x f_X(x) \, dx}{P(X < 0)}
\]
where \(f_X(x)\) is the PDF of \(X \sim \mathcal{N}(\mu, \sigma^2)\).

The PDF of \(X\) is
\[
f_X(x) = \frac{1}{\sqrt{2\pi\sigma^2}} e^{-\frac{(x - \mu)^2}{2\sigma^2}}
\]

When we condition on \(X < 0\), the PDF becomes
\[
f_{X \mid X < 0}(x) = \frac{f_X(x)}{P(X < 0)} = \frac{f_X(x)}{\Phi\left(\frac{0 - \mu}{\sigma}\right)}
\]
where \(\Phi(\cdot)\) is the CDF of the standard normal distribution.

The numerator of the conditional expectation is
\[
\int_{-\infty}^{0} e^x f_X(x) \, dx = \int_{-\infty}^{0} e^x \frac{1}{\sqrt{2\pi\sigma^2}} e^{-\frac{(x - \mu)^2}{2\sigma^2}} \, dx
\]
Combine the exponential terms:
\[
= \int_{-\infty}^{0} \frac{1}{\sqrt{2\pi\sigma^2}} e^{x - \frac{(x - \mu)^2}{2\sigma^2}} \, dx = \int_{-\infty}^{0} \frac{1}{\sqrt{2\pi\sigma^2}} e^{-\frac{x^2 - 2\mu x + \mu^2 - 2\sigma^2 x}{2\sigma^2}} \, dx
\]
Simplify the exponent:
\[
= \int_{-\infty}^{0} \frac{1}{\sqrt{2\pi\sigma^2}} e^{-\frac{x^2 - 2(\mu + \sigma^2)x + \mu^2}{2\sigma^2}} \, dx = e^{\mu + \frac{\sigma^2}{2}} \int_{-\infty}^{0} \frac{1}{\sqrt{2\pi\sigma^2}} e^{-\frac{(x - (\mu + \sigma^2))^2}{2\sigma^2}} \, dx
\]
This integral represents the CDF of a normal distribution:
\[
= e^{\mu + \frac{\sigma^2}{2}} \Phi\left(\frac{0 - (\mu + \sigma^2)}{\sigma}\right)
\]

The denominator is the probability \(P(X < 0)\):
\[
P(X < 0) = \Phi\left(\frac{0 - \mu}{\sigma}\right)
\]

\[
\mathbb{E}[e^X \mid X < 0] = \frac{e^{\mu + \frac{\sigma^2}{2}} \Phi\left(\frac{-\mu - \sigma^2}{\sigma}\right)}{\Phi\left(\frac{-\mu}{\sigma}\right)}
\]

\[
\mathbb{E}[e^X - 1 \mid X < 0] = e^{\mu + \frac{\sigma^2}{2}} \frac{\Phi\left(\frac{-\mu - \sigma^2}{\sigma}\right)}{\Phi\left(\frac{-\mu}{\sigma}\right)} - 1
\]

\[
\frac{1}{p_t \cdot \colfacN} \mathbb{E}[p_{t+1} - p_t \mid p_{t+1} < p_t] = \frac{1}{p_t \cdot \colfacN} p_t \left(e^{\mu + \frac{\sigma^2}{2}} \frac{\Phi\left(\frac{-\mu - \sigma^2}{\sigma}\right)}{\Phi\left(\frac{-\mu}{\sigma}\right)} - 1\right)
\]

\[
\frac{1}{\colfacN} \left(e^{\mu + \frac{\sigma^2}{2}} \frac{\Phi\left(\frac{-\mu - \sigma^2}{\sigma}\right)}{\Phi\left(\frac{-\mu}{\sigma}\right)} - 1\right)
\]

\end{proof}
\paragraph*{Price change expectaion}
\begin{proof}
Using the linearity of expectation:
\[
\mathbb{E}\left[\frac{p_{t+1}}{p_t} - 1\right] = \mathbb{E}\left[e^X - 1\right]
\]
This can be expanded as:
\[
\mathbb{E}\left[e^X - 1\right] = \mathbb{E}[e^X] - \mathbb{E}[1]
\]
Substituting the values from the last proof:
\[
\mathbb{E}\left[e^X - 1\right] = e^{\mu + \frac{\sigma^2}{2}} - 1
\]

\subsection{Proof of theorem \ref{theorem:equilibrium-finding}}

First we equate the borrower's utility function to zero to ensure $\E{\frac{B_{t+1}-B_t}{B_t}}=0$, this yields the equilibrium interest rate:

\begin{align*}\label{eq:equilibrium-rate}
\rS{}^* &= ( \alpha\cdot\robt + \poolDefault{}(\colfacN_t)+ \frac{\alpha}{\colfacN_t}\, \E{\indicator{\pS{t+1}-\pt<0}(\frac{\pS{t+1}-\pt}{\pt})}\\& + (1-\alpha) \cdot\E{\frac{\pS{t+1} - \pt}{\pt}} \bigg) \\&=
\alpha \,\robt + \poolDefault{}(\colfacN_t) + \frac{\alpha}{\colfacN_t} \left(e^{\mu + \frac{\sigma^2}{2}} \frac{\Phi\left(\frac{-\mu - \sigma^2}{\sigma}\right)}{\Phi\left(\frac{-\mu}{\sigma}\right)} - 1\right) + (1-\alpha) \left(e^{\mu + \frac{\sigma^2}{2}} - 1\right)
\end{align*}
If this rate is non-positive, this means that the pool has no non-trivial equilibrium in this market condition and under this $\colfacN_t$, otherwise the above rate is the unique equilibrium interest rate of the market.

Now, we assess the utility of the lender in the equilibrium interest rate:

\[\utilityl = r^* \ut- \poolDefault{}(\colfacN_t) - \rolt\]

There are two cases:

\begin{itemize}
    \item 1) $(r^*-\rolt) > \poolDefault{}(\colfacN_t) $: In this case there is a $U^* < 1$ that satisfies $\utilityl = 0$ and this utilization is $U^* = \frac{\rolt}{\rS{}^* - \poolDefault{}(\volat)}$. 
    \item 2) 1) $(r^*-\rolt) \leq \poolDefault{}(\colfacN_t) $: In this case, when the equilibrium rate is set, lenders have nonpositive utility hence they withdraw until the pool depletes and the utilization hits $1$.
\end{itemize}
\end{proof}

\subsection{Proof of theorem \ref{theorem:lse-convergence-rate}}

We borrow from the rich literature of Ordinary Least Square estimator \cite{sabourin2021lecture}. First, we note that LSE is an unbiased estimator so for any $\tau$:
\[\thetahatbvec^{\tau}(0)= \elasticityb r^* + \epsilon_0\]
 If we assume that $\elasticityb$ is known (which is a realistic assumption if $\elasticityb$ changes slower than $r^*$), $\thetahatbvec^{\tau}(0)$ also provides an unbiased estimator of $r^*$:
 \[\hat{r}^* =  r^* + \frac{\epsilon_0}{\elasticityb}\]

Now for the convergence rate, we state a special case of proposition 7 of lecture note \cite{sabourin2021lecture}:

\begin{lemma}\label{lemma:lse-lecture-note}
    Suppose that the noise is Gaussian. Denote by $\hat{\lambda}_{\tau}$ the smallest eigenvalue of  
    $\,\frac{1}{\tau}{(\pbmatt{\tau})}^T\pbmatt{\tau}$ and suppose that $\hat{\lambda}_{\tau} > 0$ for all $n \geq 1$. Then, for $k \in \{0, 1\}$, $\tau \geq 1$ and $\delta > 0$, it holds with probability $1 - \delta$,
    \begin{equation*}
        \max_{k = 0, 1} \left| \thetahatbvec^{\tau}(k) - \thetabvec^*(k) \right| \leq \sqrt{\frac{2 \zeta^2 \log(6 / \delta)}{\tau\, \hat{\lambda}_{\tau}}}.
    \end{equation*}
    Where $\thetahatbvec^{\tau}(k)$ is the $k$th element of estimated $\thetabvec$ at time $\tau$ and $\zeta$ is the noise variance.
\end{lemma}
The exploration process incorporated in Algorithm \ref{alg:interest-rate-controller} ensures the fact that there is a sub-matrix of $\,\frac{1}{\tau}{(\pbmatt{\tau})}^T\pbmatt{\tau}$ of at least size $\tau \,\nu$ with all independent rows which guarantees LSE converges.

\subsection{Proof of theorem \ref{theorem:baseline-conv-rate}}
Let's consider a piecewise linear interest rate function with one corner, for ease of analysis let's assume the operating point of the system is before the edge hence yielding a linear function $\rt = R_0 + R_s \, \ut$.

We consider the case that only the borrower is elastic hence $\rt = R_0 + R'_s \, \bt$ where $R'_s\coloneqq \frac{R_s}{\lt}$ and $\lt$ is fixed due to the lenders' in-elasticity. Now we write the borrower behavior dynamics to find the convergence rate of the protocol to $r^*$ when the market is unstable, we consider a slightly different borrower's behavior in which the absolute borrow/repay rate is proportional to the borrower's utility, even this will change the borrower's dynamic slightly but does not affect the final result, 

\begin{align*}
    \frac{\bS{t+1} - \bt}{\Deltat} &= \elasticityb(\rS{}^* - \rt)\\&= \elasticityb\left(r^* - \left(R_0 + R'_s \, \bt\right)\right)
\end{align*}
We rewrite this discrete diffenrtial equation as the general form
:
\[
f(t+1) - f(t) = af(t) + b + \epsilon_t
\]

Where $f(t)\coloneqq\bt$ and $ a\coloneqq -\Deltat\, \elasticityb\, R'_s$ and $b \coloneqq \elasticityb \, \Deltat\, (r^*-R_0)$ and $\epsilon_t$ is a guassian noise with mean $0$ and variance $\zeta^2\Deltat^2$
To find \(f(t+1)\), we rearrange the equation:
\[
f(t+1) = (1 + a)f(t) + b + \epsilon
\]

Iteratively expanding \(f(t)\) for a few steps to understand the pattern, we find:
\[
f(1) = (1 + a)f(0) + b + \epsilon_0
\]
\[
f(2) = (1 + a)f(1) + b + \epsilon_1 = (1 + a)^2 f(0) + (1 + a)b + b + (1 + a)\epsilon_0 + \epsilon_1
\]

The general form at time \(t\) can be written using a recursive expansion:
\[
f(t) = (1 + a)^t f(0) + \sum_{k=0}^{t-1} (1 + a)^k b + \sum_{k=0}^{t-1} (1 + a)^{t-1-k} \epsilon_k
\]

Simplifying the sum for \(b\), we use the geometric series sum formula:
\[
\sum_{k=0}^{t-1} (1 + a)^k = b \left( \frac{(1 + a)^t - 1}{a} \right) \quad \text{for } a \neq -1
\]

Combining all components, the complete solution to the equation is:
\[
f(t) = (1 + a)^t \left(f(0)+\frac{b}{a}\right) -\frac{b}{a} + \sum_{k=0}^{t-1} (1 + a)^{t-1-k} \epsilon_k 
\]

This function converges when $1+a=1-\Deltat\,\elasticityb\,R'_s$ is between zero and 1, in this case:
\[\E{\lim_{t \to \infty} \bt} = \frac{-b}{a} = \frac{r^*-R_0}{R'_s} \implies \E{\lim_{t \to \infty} \rt} = R_0+R_s'\E{\lim_{t \to \infty} \bt}=r^*\]

\begin{align}
    \E{\rt} &= R_0 + R'_s(1+a)^t \left(f(0)+\frac{b}{a}\right)-R_s'\frac{b}{a}\\&= R_0 +R_s'\frac{r^*-R_0}{R_s'}+ (1+a)^t\left(R_s'f(0)-(r^*-R_0)\right)\\&=r^* + (1-\Deltat\,\elasticityb\,R_s')^t\left(R_s'\bS{0}-(r^*-R_0)\right)
\end{align}

which means that in infinity, the average of the interest rate will be the same as the equilibrium interest rate, Now we assess the  variance of this process:

\[\text{Var}[\bt]=\text{Var}\left[\sum_{k=0}^{t-1} (1 + a)^{t-1-k} \epsilon_k\right]  = \sum_{k=0}^{t-1} (1 + a)^{2t-2-2k} \,\zeta^2\,\Deltat^2 = \zeta^2 \Deltat^2 \frac{1 - (1 + a)^{2t}}{1 - (1 + a)^2}\]

In order to find \convrate we need to find the variance of $\rt$:
\[\text{Var}\left[\rt\right] = {R'_s}^2 \,\text{Var}[\bt] = {R'_s}^2 \, \zeta^2 \Deltat^2 \frac{1 - (1 + a)^{2t}}{1 - (1 + a)^2}\]

We use Markov's Inequality to bound \convrate:
\[
\prob{|\rt - r^*| \geq \kappa} \leq \frac{\E{( \rt- r^*)^2}}{\kappa^2} = \frac{{R'_s}^2 \, \zeta^2 \Deltat^2 \frac{1 - (1 + a)^{2t}}{1 - (1 + a)^2}}{\kappa^2} 
\]

hence we can equate $\frac{{R'_s}^2 \, \zeta^2 \Deltat^2 \frac{1 - (1 + a)^{2t}}{1 - (1 + a)^2}}{\kappa^2} = \delta$ and write with probability more than $\delta$ we have:
\[|\rt-r^*| \leq \kappa = \sqrt{\frac{{R'_s}^2 \zeta^2 \Deltat^2 \frac{1 - (1 + a)^{2t}}{1 - (1 + a)^2}}{\delta}} = K \sqrt{\frac{1-(1-\Deltat\,\elasticityb\,R'_s)^t}{\delta}} \sim o(\frac{1}{\sqrt{\delta}})
\]

Where $K$ is a constant.

Now we consider the case where both lenders and borrowers are elastic, in this case, the equilibrium rate and utilization follow Equation \ref{eq:equilibrium-rate} and Equation \ref{eq:equilibrium-u}. This clearly shows that the equilibrium utilization is not a linear function of $r^*$, hence when setting the arbitrary $\rt = r^*$, the implied utilization by the piece-wise linear curve cannot be the equilibrium utilization of Equation \ref{eq:equilibrium-u}. this means that at that point, lenders still refer to lend more or repay and change the utilization which will change the interest rate as well.

\subsection{Proof of Lemma \ref{lemma:liq-expectation}}
First, we aim to find the liquidation $\gamma$, we know that liquidation only happens if loan to value i.e., $\frac{1}{\frac{\pS{t+1}}{p_t \cdot \colfacN}} > \liqthrshN$, hence if we define $X \coloneqq \frac{\pS{t+1}}{\pt} $, the indicator term $\indicator{X < \frac{\colfacN}{\liqthrshN}}$ will appear. Furthermore, $\gamma$ is the minimum value that satisfies $\frac{1-\gamma}{\frac{\pS{t+1}}{\pt \cdot\colfacN}-\gamma} \leq \liqthrshN$, solving for the minimum $\gamma$ will result in the term specified in the lemma statement.

We aim to find the expected value of the variable 
\[
Y = \frac{1-\frac{\liqthrshN}{\colfacN}X}{1-\liqthrshN} \indicator{X < \frac{\colfacN}{\liqthrshN}},
\]
where \( X \) is log-normally distributed with parameters \(\mu\) and \(\sigma\). This implies that \(\ln(X) \sim \mathcal{N}(\mu, \sigma^2)\).

The expected value \(E[Y]\) can be expressed as:
\[
E[Y] = E\left[\frac{1 - \frac{\liqthrshN}{\colfacN}X}{1 - \liqthrshN} \indicator{X < \frac{\colfacN}{\liqthrshN}}\right].
\]
Simplifying, we get:
\[
E[Y] = \frac{1}{1 - \liqthrshN} E\left[\left(1 - \frac{\liqthrshN}{\colfacN}X\right) \indicator{X < \frac{\colfacN}{\liqthrshN}}\right].
\]

This expectation can be written as an integral over the probability density function (pdf) of \( X \):
\[
E[Y] = \frac{1}{1 - \liqthrshN} \int_0^{\frac{\colfacN}{\liqthrshN}} \left(1 - \frac{\liqthrshN}{\colfacN}x\right) f_X(x) \, dx,
\]
where \( f_X(x) = \frac{1}{x\sigma\sqrt{2\pi}} e^{-\frac{(\ln x - \mu)^2}{2\sigma^2}} \).

Changing the variable of integration, let \( u = \ln x \). Then \( du = \frac{1}{x} dx \), so \( dx = e^u du \). The limits of integration change as follows:
- When \( x = 0 \), \( u \to -\infty \)
- When \( x = \frac{\colfacN}{\liqthrshN} \), \( u = \ln \left(\frac{\colfacN}{\liqthrshN}\right) \)

Thus, the expectation becomes:
\[
E[Y] = \frac{1}{1 - \liqthrshN} \int_{-\infty}^{\ln \left(\frac{\colfacN}{\liqthrshN}\right)} \left(1 - \frac{\liqthrshN}{\colfacN} e^u\right) \frac{1}{\sigma\sqrt{2\pi}} e^{-\frac{(u - \mu)^2}{2\sigma^2}} e^u \, du.
\]

Simplifying the integrand:
\[
E[Y] = \frac{1}{1 - \liqthrshN} \int_{-\infty}^{\ln \left(\frac{\colfacN}{\liqthrshN}\right)} \frac{1}{\sigma\sqrt{2\pi}} \left(e^{-\frac{(u - \mu)^2}{2\sigma^2}} e^u - \frac{\liqthrshN}{\colfacN} e^{-\frac{(u - \mu)^2}{2\sigma^2}} e^{2u}\right) \, du.
\]

Separate the integral:
\[
E[Y] = \frac{1}{1 - \liqthrshN} \left( \int_{-\infty}^{\ln \left(\frac{\colfacN}{\liqthrshN}\right)} \frac{e^u}{\sigma\sqrt{2\pi}} e^{-\frac{(u - \mu)^2}{2\sigma^2}} \, du - \frac{\liqthrshN}{\colfacN} \int_{-\infty}^{\ln \left(\frac{\colfacN}{\liqthrshN}\right)} \frac{e^{2u}}{\sigma\sqrt{2\pi}} e^{-\frac{(u - \mu)^2}{2\sigma^2}} \, du \right).
\]

Evaluate the integrals using the properties of the normal distribution. The first integral is:
\[
I_1 = \int_{-\infty}^{\ln \left(\frac{\colfacN}{\liqthrshN}\right)} \frac{e^u}{\sigma\sqrt{2\pi}} e^{-\frac{(u - \mu)^2}{2\sigma^2}} \, du = \Phi \left( \frac{\ln \left( \frac{\colfacN}{\liqthrshN} \right) - \mu + \sigma^2}{\sigma} \right),
\]
where \( \Phi \) is the cumulative distribution function (CDF) of the standard normal distribution.

The second integral is:
\[
I_2 = e^{\mu + \sigma^2} \Phi \left( \frac{\ln \left( \frac{\colfacN}{\liqthrshN} \right) - \mu - \sigma^2}{\sigma} \right).
\]

Combining these results, we get:
\[
E[Y] = \frac{1}{1 - \liqthrshN} \left( \Phi \left( \frac{\ln \left( \frac{\colfacN}{\liqthrshN} \right) - \mu + \sigma^2}{\sigma} \right) - \frac{\liqthrshN}{\colfacN} e^{\mu + \sigma^2} \Phi \left( \frac{\ln \left( \frac{\colfacN}{\liqthrshN} \right) - \mu - \sigma^2}{\sigma} \right) \right).
\]

\section{Notation Table} \label{app:notation-table}

\begin{table}[ht]
\centering
\begin{tabular}{|>{\raggedright}m{3cm}|>{\raggedright\arraybackslash}m{10cm}|}
\hline
\textbf{Notation} & \textbf{Description} \\
\hline
$\assetOne$ & Lent out asset (assumed to be stable coin)\\
\hline
$\assetTwo$ & Collateral assset \\
\hline
$T_m$ & Market timeframe, denoting the timeframe within which the market is stable \\
\hline
$p_t$ & Price at time $t$ \\
\hline
$\muprice$ & Drift of the price \\
\hline
$\sigma$ & Volatility of the price \\
\hline
$r_l^o$ & Risk-free lend rate offered by an external competitor \\
\hline
$r_b^o$ & Risk-free borrow rate offered by an external competitor \\
\hline
$\mathcal{P}$ & Decentralized borrowing-lending protocol \\
\hline
$r_t$ & Interest rate at time $t$ \\
\hline
$c_t$ & Collateral factor at time $t$ \\
\hline
$c$ & Collateral factor in general \\
\hline
$LT_t$ & Liquidation threshold at time $t$ \\
\hline
$LT$ & Liquidation threshold in general \\
\hline
$LI_t$ & Liquidation Incentive at time $t$ \\
\hline
$L_t$ & State variable representing total supply at time $t$ \\
\hline
$B_t$ & State variable representing total borrowings at time $t$ \\
\hline
$C_t$ & State variable representing total collateral at time $t$ \\
\hline
$U_t$ & State variable representing utilization at time $t$ \\
\hline
$\pi_{t}^i(p_{t+1})$ & Default value for borrower $i$ from timeslot $t$ to $t+1$ given that the price is $\pS{t+1}$ at time $t+1$. \\
\hline
$\pi_{t-1}(p_{t})$ & Normalized overall pool default value from timeslot $t$ to $t+1$ given that the price is $\pS{t+1}$ at time $t+1$. \\
\hline
$\pi(\colfacN_t)$ & $\E{\pi_{t}^i(p_{t+1})}$, Expectation over $\pS{t+1}$. \\
\hline
$L_{t}(i)$ & Supply of user $i$ at timeslot $t$ \\
\hline
$B_{t}(i)$ & Borrowings of user $i$ at timeslot $t$ \\
\hline
$C_{t}(i)$ & Collateral of user $i$ at timeslot $t$ \\
\hline
$\userLiq{t}{i}(\pS{t+1})$ & Liquidation incurred by user $i$ from timeslot $t$ to $t+1$, given the price to be $\pS{t+1}$ at time $t+1$ \\
\hline
$\userLiq{}{}(\colfacN_t,\liqthrshN_t)$ & $\E{\userLiq{t}{i}(\pS{t+1})}$, Expectation over $\pS{t+1}$
\\
\hline
$\utilityl$ & Utility function of a continuum lender possessing one unit of supply at time $t$ \\
\hline
$\utilitybOne$ & Utility function of a financing continuum borrower with one unit of demand at time $t$ \\
\hline
$\utilitybTwo$ & Utility function of a leveraged trading continuum borrower with one unit of demand at time $t$ \\
\hline
$\elasticityl$ & Elasticity of the lenders\\
\hline
$\elasticityb$ & Elasticity of the borrowers\\
\hline
$\alpha$ & Fraction of the financing borrowers to total\\
\hline
$\beta$ & Maximum fraction of demand/supply being controlled by adversaries\\
\hline
$\gamma$ & Default regularization factor in the optimality index definition\\
\hline
$\kappa$ & Learning rate\\
\hline
\end{tabular}
\caption{Notations and Descriptions}
\label{tab:notations}
\end{table}

\end{appendices}